\documentclass[twocolumn]{aastex631}
\usepackage{natbib}
\usepackage{graphicx}
\usepackage{url}
\newcommand{\kpc}{\,{\rm kpc}}
\newcommand{\ha}{H$\alpha$}
\newcommand{\hb}{H$\beta$} 
\newcommand{\hii}{H\,{\scriptsize II}}

\newcommand{\heii}{He\,{\scriptsize II}}
\newcommand{\ovi}{O\,{\scriptsize VI}}  
\newcommand{\oiii}{O\,{\scriptsize III}}
\newcommand{\kms}{\,km\,s$^{-1}$}  
\newcommand{\myr}{\,$M_{\odot}\,{\rm yr}^{-1}$}
\newcommand{\ro}{\,$R_{\odot}$}
\newcommand{\mo}{\,$M_{\odot}$}
\newcommand{\lo}{\,$L_{\odot}$}
\newcommand{\cmt}{\,cm$^{-3}$}
\newcommand{\cmtt}{\,cm$^{3}$}
\newcommand{\cmd}{\,cm$^{-2}$}

\newcommand{\es}{$\rm\,erg\,s^{-1}$}
\newcommand{\ecs}{$\rm\,erg\,cm^{-2}\,s^{-1}$}
\newcommand{\ecsa}{$\,\rm erg\,cm^{-2}\,s^{-1}\,\AA^{-1}$}
\def\I{\rm {\scriptsize I}}
\def\V{\rm {\scriptsize V}}
\shorttitle{Multiwavelength modeling the SED of luminous supersoft X-ray
            sources}
\shortauthors{Skopal}
\begin{document} 

\title{Multiwavelength modeling the SED of Luminous Supersoft X-ray 
       Sources in Large Magellanic Cloud and Small Magellanic Cloud}
%
\author{Augustin Skopal}
\affiliation{Astronomical Institute, Slovak Academy of Sciences, \\
             059\,60 Tatransk\'a Lomnica, Slovakia}

\begin{abstract}
Classical supersoft X-ray sources (SSSs) are understood as close 
binary systems in which a massive white dwarf (WD) accretes from 
its companion at rates sustaining a steady hydrogen burning on 
its surface generating bolometric luminosities of 
$10^{36}-2\times10^{38}$\es. 
Here, we perform for the first time the global supersoft X-rays 
to near-infrared (NIR) spectral energy distribution (SED) for 
the brightest SSSs in LMC and SMC. We test a model in which 
the ultraviolet--NIR is dominated by emission from a compact 
(unresolved) circumstellar nebula represented by the ionized gas 
out-flowing from the SSS. 
The SED models correspond to luminosities of SSSs of a few times 
$10^{38}-10^{39}$\es, radiating at blackbody temperatures of 
$\approx 3\times 10^{5}$\,K, and indicate nebular continuum, 
whose emission measure of $\gtrsim 2\times10^{60}$\cmt\ corresponds 
to a wind mass-loss at rates $\gtrsim 2\times 10^{-6}$\myr. 
Such the extreme parameters 
suggest that the brightest SSSs could be unidentified 
optical novae in a post-nova SSS state sustained at a high 
long-lasting luminosity by resumed accretion, possibly at 
super-Eddington rates. 
New observations and theoretical multiwavelength modeling of 
the global SED of SSSs are needed to reliably determine their 
parameters, and thus to understand their proper stage in stellar 
evolution. 
\end{abstract}


\section{Introduction}
\label{s:intro}
Luminous supersoft X-ray sources (SSSs) were discovered in the 
Large Magellanic Cloud (LMC) and in the Small Magellanic Cloud 
(SMC) by \cite{1981ApJ...248..925L} and \cite{1981ApJ...243..736S}. 
They are characterized by very soft thermal spectra, emitting 
predominantly at energies $\lesssim$0.5\,keV 
\citep[][]{1991A&A...246L..17G}. 
Their extremely soft spectra correspond to blackbody temperatures 
of $\approx$15--80\,eV and bolometric luminosities of 
$10^{36}-2\times10^{38}$\es. 
These objects are commonly accepted as binaries that consist 
of a massive white dwarf (WD) radiating close to the Eddington 
luminosity due to a steady nuclear burning on its surface, when 
the hydrogen-rich material is burned as fast as it is accreted, 
at rates of the order of $10^{-7}$\myr\ 
\citep[][]{1992A&A...262...97V,1997ARA&A..35...69K}. 
Following multiwavelength observations have revealed a significant 
excess of the radiation in the ultraviolet (UV) to near-infrared 
(NIR), well above the Rayleigh-Jeans tail of the SSS radiation 
itself (see Sects.~\ref{ss:0513} -- \ref{ss:0527} and \ref{s:obs}). 
To explain the high luminosity of bright SSSs, and the longer 
wavelength excess, \cite{1996LNP...472...65P} proposed a model, 
in which a significant fraction of the WD's radiation is converted 
to the UV--NIR by irradiating, (i) the flared accretion disk and 
(ii) the donor star. 

In this paper we selected four bright SSSs observed in the 
Magellanic Clouds, RX\,J0513.9-6951, RX\,J0058.6-7135, 
RX\,J0543.6-6822 and RX\,J0527.8-6954, with the aim to model 
their global X-ray---NIR spectral energy distribution (SED) 
by the method of multiwavelength modeling described by 
\cite{2015NewA...36..116S}. 
The motivation for this work is the absence of the SED model, 
which would uniformly fit radiation of SSSs in both the supersoft 
X-ray and the UV--NIR domains. 
Further, the extreme properties of SSSs and their spectral 
features indicate the presence of the nebular radiation in 
the spectrum (see Appendix~\ref{s:appA}), which could explain 
the observed strong UV--NIR excess as it is currently 
considered for the accreting nuclear-burning WDs in symbiotic 
binaries 
\citep[][]{1984ApJ...279..252K,
           1991A&A...248..458M,
           2005A&A...440..995S}. 
Therefore, here we are testing SED modeling, in which the UV--NIR 
is dominated by the nebular emission instead of the irradiated 
disk/companion radiation. 

Section~\ref{s:obs} summarizes the observed supersoft X-rays 
to NIR SED of our targets, while Sect.~\ref{s:analysis} 
introduces the method and results of our analysis. Their 
discussion is found in Sect.~\ref{s:dis}, while Sect.~\ref{s:sum} 
summarizes our findings, and proposes tasks for future 
investigation. 
In the following subsections, we first introduce our targets. 
%
%
\subsection{The LMC SSS RX\,J0513.9-6951}
\label{ss:0513}
RX\,J0513.9-6951 is a transient binary supersoft X-ray source. 
Currently, it is thought that the system contains a somewhat 
evolved star and a relatively massive compact object at a 0.76-day 
orbit, viewed nearly pole-on 
\citep[][]{1996ApJ...456..320C,1996ApJ...470.1065S}. 
The system is the brightens SSS emitting a significant amount 
of the radiation also in the far-UV ($\gtrsim 10^{-13}$\ecsa). 

RX\,J0513.9-6951 was discovered during the \textsl{ROSAT} 
All-sky Survey (1990 July -- 1991 January) in the LMC 
\citep[][]{1993A&A...270L...9S}. The authors found it to be 
variable and matched its total average spectrum by a blackbody 
radiation (see Table~\ref{tab:lit}). 
\cite{1993A&A...278L..39P} revealed the optical counterpart as 
a $B\approx 17$\,mag blue star. Its spectrum was dominated by 
the hydrogen Balmer lines with a signature of a broad P-Cyg 
absorption component indicating a mass-loss from the system 
at $v_{\infty} \approx 3600$\kms. Absence of He\,\I\ lines 
and the presence of highly ionized ions (O\,\V\I\,3811, 3834 
doublet, He\,\I\I) with 
$I_{\rm{He\,\I\I}\,4686} \gg I_{\rm{H}\beta}$ reflect 
the presence of a very hot ionizing source in the binary. 
They also pointed that the optical counterpart appears 
fainter during the X-ray outburst. 

Pronounced optical variability of RX\,J0513.9-6951 was revealed 
by optical multicolor monitoring campaigns. 
\cite{1996A&A...309L..11R} measured an optical low states 
during February to December, 1993, lasting around 40 days with 
a decrease in the continuum by $\Delta V \sim 1$\,mag and 
a factor of $\approx$10 decrease in emission line fluxes. 
The light curve obtained within the MACHO project specified 
the occurrence of the low stages on a time-scale of 
$\sim 100-200$ days 
\citep[][]{1996MNRAS.280L..49A,1996ApJ...470.1065S}. 
In addition, during the high-state, these authors revealed 
a small ($\Delta V \sim 0.05$) light modulation with a period 
of 0.763\,d, which is consistent with variations in the radial 
velocities of \heii\,$\lambda$4686 emission line, and thus 
confirms a binary nature of the system 
\citep[][]{1996ApJ...456..320C}. 

Broad wings of the strongest emission lines (He\,\I\I\ and \hb) 
accompanied with satellite components indicated a bipolar outflow 
and thus the presence of a disk in the binary. The bipolar 
outflow in the form of jets suggested that the accretor is 
a WD \citep[][]{1996ApJ...470.1065S,1996ApJ...456..320C}. 

Based on all previously published optical and X-ray data, 
\cite{1996ApJ...470.1065S} pointed out the principal feature of 
the variability: the anti-correlation between the supersoft 
X-ray and the optical fluxes. 
X-ray monitoring of RX\,J0513.9-6951 showed that the flux 
anti-correlation is very strict: There are cyclic changes 
between optical-low/X-ray-on states and optical-high/X-ray-off 
states \citep[e.g.,][]{2000A&A...354L..37R,2010NewAR..54...75C}. 

Based on the \textsl{HST} ultraviolet spectroscopy, 
\cite{1998A&A...333..163G} derived 
the neutral hydrogen density 
$N_{\rm H} = (5.5\pm 1.0) \times 10^{20}$\cmt. 
The authors found that this value is consistent with parameters 
determined by model atmosphere fits to the \textsl{ROSAT PSPC} 
spectrum of July 1993 (see Table~\ref{tab:lit}). 

\subsection{The SMC SSS RX\,J0058.6-7135 (LIN~333)}
\label{ss:0058}
\cite{1956ApJS....2..315H} labeled this object in his 
``Catalog of Emission Nebulae in the Small Cloud'' as 
LH$\alpha$-115~N67. Independently, \cite{1961AJ.....66..169L} 
put it in his catalog 
under the number 333, remarking features of a planetary nebula in 
its spectrum. Therefore, the identifier LIN~333 is frequently 
used for this object. \cite{1987ApJ...320..159A} analyzed its 
\textsl{IUE} spectra, and revealed a strong nebular emission 
produced by the object, characterized with the emission measure 
of $\sim 1.4\times 10^{60}$\cmt\ (see Appendix~\ref{s:appA}) 
and a high electron temperature of $\sim 30\,000$\,K. 
They also measured the optical position of the nebula, which 
helped \cite{1991MNRAS.252P..47W} to identify 
LH$\alpha$-115~N67 with the strong SSS 1E0056.8-7154, which 
is also named as SMP~SMC 22 \citep[][]{1978PASP...90..621S}. 
\cite{1991MNRAS.252P..47W} suggested that 1E0056.8-7154 
is the single hot nucleus of the planetary nebula 
LH$\alpha$-115~N67. 
The X-ray detection of this source was first reported by 
\cite{1981ApJ...243..736S} on the basis of observations with 
the \textsl{Einstein} satellite. 
Following X-ray observations were carried out by the 
\textsl{ROSAT} and \textsl{XMM-Newton} satellites, see 
\cite{1994A&A...288..538K} and \cite{2010A&A...519A..42M}, 
respectively. 
Modeling the X-ray SED of LIN~333 was performed by 
\cite{1994A&A...288..538K}, \cite{1994A&A...288L..45H}, 
and \cite{2010A&A...519A..42M}. Corresponding parameters 
are introduced in Table~\ref{tab:lit}. 
Based on their models, \cite{1994A&A...288L..45H} found 
that accreting WD in a binary is consistent with the 
\textsl{ROSAT} spectrum of 1E0056.8-7154, but also admitted 
a possibility that the SSS is an exceptionally hot single 
central star of a planetary nebula. 
Comparing all X-ray data and their models, spanning almost 
30 years, \cite{2010A&A...519A..42M} did not find any apparent 
long-term variability in the luminosity or spectrum of LIN~333. 
The authors pointed out that the lack of variability, typical 
for other SSSs, and too low upper limit to the optical 
counterpart \citep[$V\sim 21$,][]{2004ApJ...614..716V} do not 
favor a binary nature of LIN~333. Therefore, they supported 
the interpretation that it is a single, very hot star on its 
way to becoming a relatively massive ($\sim$1\mo) WD. 
Using the \textsl{Spitzer Space Telescope} infrared spectra, 
\cite{2009ApJ...699.1541B} classified SMP~SMC~22 as 
a high-excitation planetary nebula with high excitation 
lines ([O\,{\small IV}] and [Ne\,{\small V}]). The absence of 
dust-related features in the spectrum was ascribed to the high 
temperature and luminosity of SMP~SMC~22. 

In spite of these arguments, we interpret the results of our 
multiwavelength modeling of the LIN~333 SED within a binary-star 
model, in which a WD accretes from its companion at a high 
rate (Sect.~\ref{ss:lum}). The main reason for this assumption is 
the exceptionally high luminosity of LIN~333 compared to other 
planetary nebulae, indication of a compact unresolved nebula 
in its spectrum -- common main features of all our targets 
(Fig.~\ref{fig:seds}, Sect.~\ref{ss:ducn}, Appendix~\ref{s:appA}), 
and possible presence of the binary companion indicated by 
the optical fluxes (see Appendix~\ref{s:appC}). 
%
%
%
\begin{table*}[t]
\caption{Physical parameters of our targets from the 
literature (designation as in Sect.~\ref{ss:sed})}
\label{tab:lit}
\begin{center}
\begin{tabular}{rcccccc}
\hline
\hline
\noalign{\smallskip}
Object                          &
$N_{\rm H}$                     & 
$R_{\rm SSS}^{\rm eff}$         & 
$T_{\rm BB}/T_{\rm eff}$        & 
$\log(L_{\rm SSS})$             & 
Model                           &
Ref.                            \\
                                & 
[$10^{20}\,{\rm cm^{-2}}$]      &
[ $R_{\odot}$ ]                 &
[ K ]                          &
[ ${\rm erg\,s^{-1}}$ ]         &
BB/Atm.                         &
                                \\
%
%
\noalign{\smallskip}
\hline
\noalign{\smallskip}
RX\,J0513 & 9.4 & $\sim$0.039 & 456000$\pm 78000$ & $\sim$38.37 
          & BB$^{a}$ & 1 \\
RX\,J0513 & 4--7  & 0.009--0.017 & $\approx$560000 & 37.40--37.95 
          & Atm.$^{b}$ & 2 \\
LIN~333   & 5.2 - 16   & 0.086   & 160000 - 400000 &  38.28
          & BB$^{l}$   & 9 \\
LIN~333   & 5   & 0.012       & 440000      &       37.30 
          & Atm.$^{c}$ & 3 \\
LIN~333   & 8   & 0.072       & 360000      &       38.48
          & BB$^{c}$   & 3 \\
LIN~333   & 2.8--3.2 & 0.19--0.22 & 154000  &  37.81--37.93
          & Atm.$^{d}$ & 4 \\
LIN~333   & 5.2      & 0.049      & 313000  &       37.91
          & BB$^{d}$ & 4 \\
CAL~83    & 17       & $>0.06$ & 302000 & $L_{\rm SSS}> L_{\rm Edd}$ 
          & BB$^{e}$ & 5 \\
CAL~83    & 8        & 0.022   & 500000 & $L_{\rm SSS}\equiv L_{\rm Edd}$
          & BB$^{f}$ & 5 \\
CAL~83    & 8.87     & 0.014--0.016 & 516000 &       37.76
          & Atm.$^{g}$ & 6 \\
CAL~83    & 6.07     & $\gtrsim$0.015 & $\lesssim$403000 & $\gtrsim$37.30
          & Atm.$^{g}$ & 6 \\
CAL~83    & 6.5      & 0.012$\pm 0.003$  & 539000$\pm 16000$ & 37.40$\pm 0.03$
          & BB$^{h}$ & 7 \\
CAL~83    & 6.5      & 0.018$\pm 0.001$  & 378000$\pm 9000$ & 37.37$\pm 0.01$
          & Atm.$^{h}$ & 7 \\
CAL~83    & 6.5$\pm 1.0$ & 0.01$\pm 0.001$ & 550000$\pm 25000$ & 37.54$\pm 0.13$
          & Atm.$^{i}$ & 8 \\
RX\,J0527 & 14       & $>0.16$  & 209000 & $L_{\rm SSS}> L_{\rm Edd}$ 
          & BB$^{j}$ & 5 \\
RX\,J0527 & 8.5      & 0.043    & 348000 & $L_{\rm SSS}\equiv L_{\rm Edd}$ 
          & BB$^{k}$ & 5 \\
\noalign{\smallskip}
\hline
\end{tabular}
\end{center}
$^{a}$ \textsl{ROSAT}, October 30, 1990; 
1 - \cite{1993A&A...270L...9S}; 
$^{b}$ \textsl{ROSAT}, July 1993; 
2 - \cite{1998A&A...333..163G}; 
$^{c}$ \textsl{ROSAT}, October 8, 1991; 
3 - \cite{1994A&A...288L..45H}; 
$^{d}$ \textsl{XMM-Newton}, dates in Table~\ref{tab:log}; 
4 - \cite{2010A&A...519A..42M}; 
$^{e}$ \textsl{ROSAT}, June 20, 1990 - best fit; 
5 - \cite{1991A&A...246L..17G}; 
$^{f}$ as in $^{e}$, but for the Eddington luminosity ($L_{\rm Edd}$); 
$^{g}$ \textsl{ROSAT}, observation No. 500131p; 
6 - \cite{1998A&A...331..328K}; 
$^{h}$ \textsl{BeppoSAX}, $N_{\rm H}$ fixed, March 7-8, 1997; 
7 - \cite{1998A&A...332..199P}; 
$^{i}$ \textsl{Chandra, XMM-Newton}, $N_{\rm H}$ fixed, 
       dates in Table~\ref{tab:log}; 
8 - \cite{2005ApJ...619..517L}; 
$^{j}$ \textsl{ROSAT}, June 16-25, 1990 - best fit; 
$^{k}$ as in $^{j}$, but for $L_{\rm Edd}$; 
$^{l}$ \textsl{ROSAT}, October 27-31, 1990; 
9 - \cite{1994A&A...288..538K}. 
\normalsize
\end{table*}
\subsection{The LMC SSS RX\,J0543.6-6822 (CAL~83)}
\label{ss:cal83}
The bright SSS CAL~83 was discovered in the {\em Einstein} 
survey of the LMC \citep[][]{1981ApJ...248..925L}. Subsequent 
observations with \textsl{EXOSAT} from November 1983 confirmed 
its very soft X-ray spectrum with an upper limit of 0.3 count/s 
\citep[][]{1985SSRv...40..229P}. The optical spectrum from May 1984 
revealed its similarity to that of the low mass X-ray binaries 
LMC~X-2 and Sco~X-1. From radial velocity variations of the 
optical emission lines (H and \heii\,$\lambda$4686), 
\cite{1987ApJ...321..745C} discovered that CAL~83 is a binary 
system with a period of $\sim 1$ day. Orbital elements from radial 
velocities of the \heii\,$\lambda$4686 line were improved by 
\cite{2004AJ....127..469S}. 

Following X-ray observations of CAL~83 were carried out with
the \textsl{ROSAT}, \textsl{BeppoSAX}, \textsl{Chandra} and 
\textsl{XMM-Newton} satellites. 
\cite{1998A&A...331..328K} presented a 1.5-year 
X-ray light curve of CAL~83 as measured with the \textsl{ROSAT/HRI} 
from July 1995 to December 1996. 
The authors also indicated the first short-term X-ray-off state 
of CAL~83 on JD~2\,450\,200 (April 26, 1996). The second and 
the third  X-ray-off state was reported by 
\cite{2002A&A...387..944G} and \cite{2005ApJ...619..517L} on 
the basis of \textsl{Chandra} observations from November 1999 
and October 2001, respectively. This demonstrates recurrent nature 
for CAL~83. 
Modeling the X-ray SED of CAL~83 was carried out by 
\cite{1991A&A...246L..17G}, 
\cite{1998A&A...331..328K}, 
\cite{1998A&A...332..199P}, 
\cite{2001A&A...365L.308P} and 
\cite{2005ApJ...619..517L}. 
Corresponding parameters are found in Table~\ref{tab:lit}. 

The prototypical SSS CAL~83 was also intensively studied 
in the ultraviolet and optical. It was identified with 
a variable, blue, $V \approx 17$, point like source with an 
orbital period of 1.047 days
\citep[][]{1984ApJ...286..196C,
1985SSRv...40..229P,
1988MNRAS.233...51S,
2004AJ....127..469S,
2013MNRAS.432.2886R}. 
Optical 3900--6800\,\AA\ spectrum from May 1984 showed relatively 
weak Balmer emission lines and a high \heii\,$\lambda$4686/\hb\ 
ratio ($> 5$). The nebular [O\I\I\I]\,5007\,\AA\ line was also 
present. The blue-shifted \heii\,$\lambda$4686 emission component 
suggested a mass outflow with velocities at 1500--2300\,\kms\
\citep[][]{1985SSRv...40..229P}. 
CAL~83 is known to show irregular brightness variations
of more than a magnitude over timescales of months. Using 
the MACHO $V$ and $R$ light curves (1993--1999), 
\cite{2002A&A...387..944G} indicated that CAL~83 exhibits 
distinct and well-defined low, intermediate, and high optical 
states, characterized with no color changes during transition 
phases. They also found that X-ray-off states were observed 
during optical high states. A variability has also been found 
in the ultraviolet by \cite{1996LNP...472..107G}, who observed 
CAL~83 during high and low states. 
More recent optical and X-ray observations revealed 
a quasi-periodic variation on a time-scale of 450\,d with 
the optical low and high states lasting for $\sim$200 and 
$\sim$250\,d \citep[][]{2013MNRAS.432.2886R}. The authors also 
clearly established the X-ray/optical flux anticorrelation. 
It was found that the X-ray light curve exhibits significant 
modulations with periodicities from 67\,s to hours
\citep[][]{2014MNRAS.437.2948O,2017MNRAS.467.2797O}. 
It is generally assumed that variations in the optical flux 
are most likely associated with changes in the accretion disk 
\citep[e.g.,][]{2002A&A...387..944G}. 

Optical imaging of CAL~83 revealed the presence of an extended 
ionized nebula surrounding CAL~83, first recognized by 
\cite{1989ESOC...32..285P} on the [\oiii]\,$\lambda$5007 CCD 
image with a diameter of $\sim50\arcsec$. 
\cite{1995ApJ...439..646R} reported that the nebula is caused by 
the illumination of the local ISM by the central SSS CAL~83 with 
the time-averaged X-ray luminosity $>3\times10^{37}$\es. 
The authors inferred that the nebula has a bright inner zone with 
a characteristic radius, particle density and mass of $\sim$7.5\,pc, 
4--10\,cm$^{-3}$ and $\gtrsim$150\mo, respectively, while the 
diffuse outer nebula has a characteristic radius of $\sim$20\,pc. 
Based on the spectra taken over a $25\arcsec\times25\arcsec$ 
field of view (FoV, i.e., $7.5\times7.5$\,pc$^2$ for the distance 
of 55\,kpc), \cite{2012A&A...544A..86G} 
presented the morphology of the ionized gas within the inner 
nebula, for the first time indicated there a \heii\,$\lambda$4686 
emission region, and extracted the global spectrum of the ionization 
nebula within their FoV (see Appendix~\ref{s:appB}, in detail). 
Using the {\small CLOUDY} software and typical parameters of 
a SSS, the authors modeled the CAL~83 nebula. However, none of 
their models was able to fit the observed spectrum very well. 
   It is of interest to note that searching for ionization nebulae 
surrounding other strong SSSs in LMC and SMC has been negative, 
which indicates that either the ISM densities in their environments 
must be significantly lower than surrounding CAL~83, or the average 
X-ray luminosities of these sources over the last $\lesssim$10000 
years must be significantly lower than presently observed 
\citep[][]{2016MNRAS.455.1770W,2020MNRAS.497.3234F}. 
\subsection{The LMC SSS RX\,J0527.8-6954}
\label{ss:0527}
This source was discovered during the calibration phase of 
\textsl{ROSAT} pointing the LMC field between 16 and 25 June 
1990 \citep[][]{1991A&A...246L..17G}. Profile of its supersoft 
spectrum was very similar to that of CAL~83, but it was fainter 
by a factor of $\sim 10$. The model SED was done by 
\cite{1991A&A...246L..17G} (Table~\ref{tab:lit}). 

Repeated observations of this source exhibited a continuous 
decline in the photon counts, from $\sim 0.2$ to 
$\sim 0.01$\,cts\,s$^{-1}$, 
during 1990 to 1995 \citep[][]{1996A&A...312...88G}. 
The authors speculated that a likely mechanism causing the 
X-ray variability could be a hydrogen flash on a massive WD 
($\ge 1.1$\mo) accreting at a high rate of $10^{-7}$\myr. 
This view is supported by the fact that the source was not 
detected by the {\em Einstein} satellite 20 years ago, 
which thus appears that the source is recurrent and was in 
a declining phase during the \textsl{ROSAT} observations. 
Also the 4-yr light curve revealed a steady fading by 
$\lesssim$0.5\,mag and a 0.39-day modulation with 
$\sim$0.05\,mag semi-amplitude. 

Based on the optical photometry obtained via the MACHO project, 
\cite{1997MNRAS.291L..13A} first suggested an optical counterpart 
to RX\,J0527.8-6954, and \cite{1997PASP..109...21C} specified it 
as a B8\,IV star with $V = 17.3$\,mag. Using a high-resolution 
imaging, \cite{2010A&A...517L...5O} identified the X-ray source 
with a B5e\,V star that is associated with a bipolar \ha\ 
subarcsecond extended emission. The authors pointed that for 
a massive WD and the Be\,V companion with $M \sim 6$\mo, the 
orbital period has to exceed 21 hours, and thus the 0.39-day 
period cannot be associated with the orbit. 
%
%
%
\begin{table*}
\caption{Log of the used observations from the space}
\label{tab:log}
\begin{center}
\begin{tabular}{ccccc}
\hline
\hline
\noalign{\smallskip}
    Date   & Julian date  & Region  & Observing   & $T_{\rm exp}$ \\
yyyy/mm/dd & JD~2\,4...   &  [nm]   & satellite   &      [ks]     \\
\noalign{\smallskip}
\hline
\noalign{\smallskip}
\multicolumn{5}{c}{RX\,J0513.9-6951} \\
1990/10/30$^{1}$ & 48195.5 & 1.92--6.2 & 
                       \textsl{ROSAT}$^{2}$ & 1.9 \\
2001/10/31 & 52214.5   & 94--118  &\textsl{FUSE}$^{3}$ & 6.6 \\
2003/07/05 & 52827.0   & 94--118  &\textsl{FUSE}$^{3}$ &33.0 \\
1996/11/13 & 50401.7   &115--144 &\textsl{HST}$^{3}$   & 2.8 \\
1992/04/24 & 48736.6   &115--198 &\textsl{IUE}$^{3}$   &24.0 \\
1992/04/25 & 48737.6   &186--335 &\textsl{IUE}$^{3}$   &23.0 \\
1995/06/29 & 49898.4   &115--198 &\textsl{IUE}$^{4}$   &22.8 \\
\multicolumn{5}{c}{RX\,J0058.6-7135 (LIN~333)}               \\
1991/10/08 & 48537.5   &2.5--6.2  &\textsl{ROSAT}      &17   \\
2007/06/04 & 54256.8   &2.5--6.3  &\textsl{XMM-Newton} & 7.7 \\
2009/05/14 & 54966.3   &2.5--6.3  &\textsl{XMM-Newton} &16.54\\
2009/09/25 & 55100.3   &2.5--6.3  &\textsl{XMM-Newton} &32.3 \\
2002/07/25 & 52480.5   & 92--119  &\textsl{FUSE}       &26.4 \\
1984/04/16 & 45806.5   &150 - 190 &\textsl{IUE}        &14   \\  
1984/04/17 & 45807.5   &190 - 333 &\textsl{IUE}        &16   \\  
1995/03/20 & 49796.5   &150 - 540 &\textsl{HST}        & 2.5 \\ 
\multicolumn{5}{c}{RX\,J0543.6-6822 (CAL~83)} \\
1990/06/20 & 48063.5 & 2.1--5.9   &\textsl{ROSAT}      & 3.6 \\
2000/04/23 & 51658.0 & 2.4--5.4   &\textsl{XMM-Newton} &45.1 \\
2001/08/15 & 52137.3 & 2.4--5.4   &\textsl{Chandra}    &35.4 \\
2001/09/22 & 52175.3 & 93.5--118  &\textsl{FUSE}       &45.6 \\
1996/11/10 & 50398.0 &  115--144  &\textsl{HST}        & 2.8 \\
1984/11/24 & 46029.1 &  115--198  &\textsl{IUE}        &15.1 \\ 
1987/12/12 & 47142.0 &  115--198  &\textsl{IUE}        &14.4 \\
1990/10/29 & 48194.1 &  115--198  &\textsl{IUE}        &24.4 \\
1990/11/07 & 48203.4 &  115--198  &\textsl{IUE}        &22.8 \\
1987/12/11 & 47141.0 &  186--332  &\textsl{IUE}        &16.1 \\
1988/08/12 & 47397.3 &  197--333  &\textsl{IUE}        &12.9 \\
%
\multicolumn{5}{c}{RX\,J0527.8-6954} \\
1990/06/20 & 48063.5 & 2.6--5.9   &\textsl{ROSAT}      &18.8 \\
1992/09/01 & 48867.2 & 115--198   &\textsl{IUE}        &16.7 \\
1992/09/01 & 48867.3 & 240--318   &\textsl{IUE}        & 6.0 \\
\noalign{\smallskip}
\hline
\end{tabular}
\end{center}
  $^{1}$~avearage of the second part, 
  $^{2}$~the total spectrum 
         \citep[see Fig.~2 of][]{1993A&A...270L...9S}, 
  $^{3}$~high state, $^{4}$~low state. 
\end{table*}
\section{Observations}
\label{s:obs}
This section summarizes observations of our targets we used 
to model their global SED in the continuum. 
Spectroscopic observations with {\em Far Ultraviolet Spectroscopic 
Explorer} (\textsl{FUSE}), {\em Hubble Space Telescope} 
(\textsl{HST}) and {\em International Ultraviolet Explorer} 
(\textsl{IUE}) were obtained from the satellite archives with 
the aid of the MAST. 
Other observations were taken from the literature and catalogs 
as referred in the following subsections. Data obtained from 
the space are collected in Table~\ref{tab:log}. 
Stellar magnitudes in the optical--NIR were converted to fluxes 
according to the calibration of \cite{1982asph.book.....H} 
and \cite{1979PASP...91..589B}. 
To correct the observed X-ray fluxes for absorptions we used the 
{\em tbabs} absorption model for the interstellar matter 
composition with abundances given by \cite{2000ApJ...542..914W} 
(e.g., $\log(A_{\rm OI})+12 = 8.69$). 
Observations for $\lambda > 912$\,\AA\ were dereddened with 
$E_{\rm B-V}$ = 0.06 and 0.08 (unless otherwise specified), 
and physical parameters were scaled with the distance 
$d$ = 49 and 60\,kpc to objects in the LMC and SMC, respectively
\citep[][]{1998ARA&A..36..435M}. 
%
\subsection{RX\,J0513.9-6951} 
\label{ss:0513obs}
The observed SED of RX\,J0513.9-6951, used in this contribution, 
covers the spectral range from 1.92\,nm to 2200\,nm. It consists 
of the \textsl{ROSAT PSPC} observations (1.92 -- 6.20\,nm) as 
published by \cite{1993A&A...270L...9S}. The data were adopted 
from their figure 2. 
The far-UV spectra with \textsl{FUSE} (B0160104, D0060102) 
were taken during the high state of RX\,J0513.9-6951 as found 
by \cite{2002AJ....124.2833H} for the 2001 observation, and 
according to the light curve of \cite{2008A&A...481..193B} for 
the 2003 spectrum. 
The average continuum level of both spectra is comparable, 
with $F(2001)/F(2003) \sim 1.2$. 
According to the light curve of \cite{2002AJ....124.2233C}, 
the high resolution \textsl{HST/GHRS} spectrum (Z3HS0205T) 
was also taken during a high state. 
Further spectroscopic observation in the ultraviolet was 
carried out with the \textsl{IUE} satellite. The 1992 spectrum 
(SWP44467, LWP22883) corresponds to a high state, because 
its short-wavelength part fits well the \textsl{HST} spectrum. 
However, the 1995 spectrum (SWP55170) was taken at the optical 
minimum, i.e., during a low state \citep[][]{2002AJ....124.2233C}. 

Observations from the space were supplemented with the
ground-based optical spectroscopy obtained during both 
the high state \citep[][]{1993A&A...278L..39P} and the low 
state \citep[][]{1996A&A...309L..11R}. 
The former was taken nearly simultaneously with the \textsl{IUE} 
high-state spectra from 1992. In addition, optical $BVR$ 
broad-band photometry was estimated from Fig.~1 
of \cite{1996A&A...309L..11R}. Further $RI$ and infrared 
$JHK$ photometric measurements were found in the USNO-B1.0 
Catalog (R = 16.54, I = 15.75) \citep[][]{2003AJ....125..984M}, 
the NOMAD Catalog (V = 16.640, R = 16.540) 
\citep[][]{2005yCat.1297....0Z}, and the 2MASS All-Sky Catalog 
of Point Sources ($J = 16.492\pm 0.157, H \sim 15.517, 
K\sim 15.728$) \citep[][]{2006AJ....131.1163S}. 
\subsection{RX\,J0058.6-7135 (LIN~333)}
\label{ss:0058obs}
Multiwavelength observations, we used to model the SED of this 
SSS, cover the spectral range from the supersoft X-rays to 
the optical ($\sim 2.4-360$\,nm). The supersoft X-ray fluxes 
made by the \textsl{ROSAT} satellite were reconstructed with 
the aid of Fig.~2 in \cite{1994A&A...288L..45H}, and those 
carried out with the \textsl{XMM-Newton} satellite were estimated 
according to the blackbody model of \cite{2010A&A...519A..42M}. 
The ultraviolet spectra were obtained from the satellite 
archives of \textsl{FUSE} (C0560301000), \textsl{IUE} 
(SWP22766, LWP3168) and \textsl{HST FOS} (Y2N30104T, Y2N30105T, 
Y2N30106T). 
Spectroscopic observations were supplemented with the 
$UBVR$ broad-band photometry, taken from the catalogs of 
\cite{2002ApJS..141...81M} (U = 16.16, B = 16.86, V = 16.37, 
R = 16.09), \cite{2002AJ....123..855Z} (U = 16.643, B = 17.240, 
V = 16.924), \cite{2010yCat....102023S} (V = 16.90), and the 3rd 
release of the \textsl{DENIS} catalog (I = 17.662, J = 16.156). 
\subsection{RX\,J0543.6-6822 (CAL\,83)}
\label{ss:cal83obs}
%
The observed SED of CAL\,83, used in this paper, covers 
the spectral range from 2.07\,nm to 790\,nm. 
The calibrated supersoft X-ray fluxes were reconstructed 
from Fig.~1 of \cite{1991A&A...246L..17G} (\textsl{ROSAT}) 
and Fig.~1 of \cite{2005ApJ...619..517L} 
(\textsl{Chandra, XMM-Newton}). 
In both cases the source was in the X-ray-on state 
\citep[see][for the ROSAT observation]{1998A&A...331..328K}. 

In the ultraviolet, eleven far-UV spectra of CAL\,83 were 
carried out by \textsl{FUSE} (B0150101) on September 22, 2001 
\citep[][]{2004AJ....127..469S}. The continuum fluxes were estimated 
within the range of 93.5 -- 118\,nm. 
The high resolution \textsl{HST/GHRS} spectrum (Z3HS0105T) covers 
the region from $\sim 114.9$ to $\sim 143.5$\,nm. 
Further spectroscopic observation in the ultraviolet was 
carried out with the \textsl{IUE} satellite. Continuum of 
the SWP spectra No. 24554, 32511, 39995, 40075 and the LWP 
spectra No. 12257, 13902 were consistent best with those 
of the \textsl{HST} and \textsl{FUSE} spectra. Therefore, 
we used their average fluxes weighted with exposure times 
to model the SED. Other \textsl{IUE} spectra had a higher 
continuum, in particular the SWP30024 spectrum was a factor 
of $>2$ above the used minimum values. 
According to the measured variations in both the X-ray 
and the UV--optical fluxes and their anticorrelation 
\citep[see][]{2002A&A...387..944G}, it is probable that 
the minimum level of the UV continuum, which we used to model 
the SED, corresponds to the X-ray-on phase of CAL\,83. 

Spectroscopic observations were supplemented with the optical 
$UBVRI$ broad-band photometry, taken from catalogs of 
\cite{2002ApJS..141...81M} ($U = 16.25$, $B = 17.18$, $V = 17.44$), 
\cite{2004AJ....128.1606Z} ($U=16.022$, $B=16.864$, $V=16.976$) 
and the 3rd release of the \textsl{DENIS} database 
($R = 17.0$, $I = 16.538$). Compared are values reported 
by \cite{1987ApJ...321..745C} ($B = 17.19-17.34$, $V=17.22-17.35$), 
and fluxes from the optical, 400--680\,nm, spectrum
published by \cite{1985SSRv...40..229P} 
scaled with their average photometric measurements 
($U-B = -1.04\pm 0.06$, $B-V = 0.02\pm 0.06$, $B = 17.2\pm 0.3$). 
%
In the near-IR \cite{2007PASJ...59..615K} measured $J = 16.98$, 
$H = 16.81$ and $K = 16.77$, and 2MASS 6X Point Source Working 
Database/Catalog reports $J = 17.433$, $H = 17.016$ and 
$K = 16.079$. 
%
%
\begin{table*}[t]
\scriptsize
\caption{Physical parameters of bright SSSs in LMC and SMC from 
         multiwavelength SED models (see Sect.~\ref{ss:sed})}
\label{tab:par}
\begin{center}
\begin{tabular}{c|ccccc|cc|c}
\hline
\hline
\noalign{\smallskip}
 Object                         &
\multicolumn{5}{c|}{Supersoft X-ray source}   &
\multicolumn{2}{c|}{Nebula}     &
                                \\
                                &
$N_{\rm H}$                     & 
$R_{\rm SSS}^{\rm eff}$         & 
$T_{\rm BB}$                    & 
$\log(L_{\rm SSS})$             & 
$\log(L_{\rm X})^{7}$           &
$T_{\rm e}$                     & 
$EM$                            & 
$\chi^2_{\rm red}$/d.o.f.       \\
                                & 
[$10^{20}\,{\rm cm^{-2}}$]      &
[$R_{\odot}$]                 &
[K]                           &
[${\rm erg\,s^{-1}}$]         &
[${\rm erg\,s^{-1}}$]         &
[K]                           & 
[$10^{60}\,{\rm cm^{-3}}$]    & 
                                \\
%
%
\noalign{\smallskip}
\hline
\noalign{\smallskip}
RX\,J0513$^{1}$~ &
        12$\pm 2$ & 0.19$\pm 0.02$~&~$350000\pm 10000$~ & 39.28$\pm 0.14$~&
        38.25$\pm 0.13$ & 30000$\pm 10000$ & 2.3$\pm 0.2$ & 2.8 / 40  \\
RX\,J0513$^{2}$~ &
        -- &0.26$\pm 0.03$ & $350000^{3}$ & 39.54$\pm 0.15$ & 
        38.52$\pm 0.14$ & 30000$\pm 10000$ & 8.6$\pm 0.5$ & 3.6 / 38  \\
LIN~333$^{4}$~ &
        9.0$\pm 0.2$ & 0.17$\pm 0.02$ & $305000\pm 10000$ & 38.96$\pm 0.13$ &
        37.66$\pm 0.12$ & 37000$\pm 5000$ & 1.6$\pm 0.2$ & 2.6 / 39  \\
LIN~333$^{5}$~ &
        7.1$\pm 0.2$ & 0.18$\pm 0.02$ & $267000\pm 7000$ & 38.75$\pm 0.12$ &
        37.45$\pm 0.11$ & --          & --         & 4.4 / 22  \\
CAL~83$^{4}$~ &
        10.3$\pm 0.5$&0.15$\pm 0.02$ & $345000\pm 15000$ &39.04$\pm 0.16$ &
        37.99$\pm 0.15$ & 30000$\pm 15000$ & 2.9$\pm 0.3$ & 3.8 / 61  \\
CAL~83$^{6}$~ &
        12.6$\pm 0.5$&0.16$\pm 0.02$ & $357000\pm 15000$ &39.14$\pm 0.16$ &
        38.15$\pm 0.15$ & --           & --         & 2.9 / 35  \\
RX\,J0527~ &
        7.1$\pm 0.2$&0.16$\pm 0.02$ & $255000\pm 20000$ &38.57$\pm 0.06$ &
        36.83$\pm 0.05$ & 9000$\pm 2000$ & 4.3$\pm 0.3$ & 3.3 / 34  \\
\noalign{\smallskip}
\hline
\end{tabular}
\end{center}
$^{1}$ low state, 
$^{2}$ high state: the $(1-9)\times 10^3$\,\AA\ model SED for 
        $T_{\rm h}$ adapted from the low state 
        (Fig.~\ref{fig:rx0513sed2}), 
$^{3}$ fixed value, 
$^{4}$ model SED with the \textsl{ROSAT} data, 
$^{5}$ model SED with the \textsl{XMM-Newton} data, 
$^{6}$ model SED with the \textsl{Chandra/XMM-Newton} data, 
$^{7}$ the luminosity in the 0.2--1.0\,keV range. 
\normalsize
\end{table*}
\subsection{RX\,J0527.8-6954}
\label{ss:0527obs}
In this case, the available observed SED in the continuum 
covers the spectral range from 2.65\,nm to 550\,nm. 
The calibrated supersoft X-ray fluxes were obtained by
the \textsl{ROSAT} satellite during the pointing observations
of the LMC field in 1990 \citep[][]{1991A&A...246L..17G}. 
According to \cite{1996A&A...312...88G} the source was at 
the brighter stage. 

\begin{figure*}[t]
 \begin{center}
\resizebox{\hsize}{!}
          {\includegraphics[angle=-90]{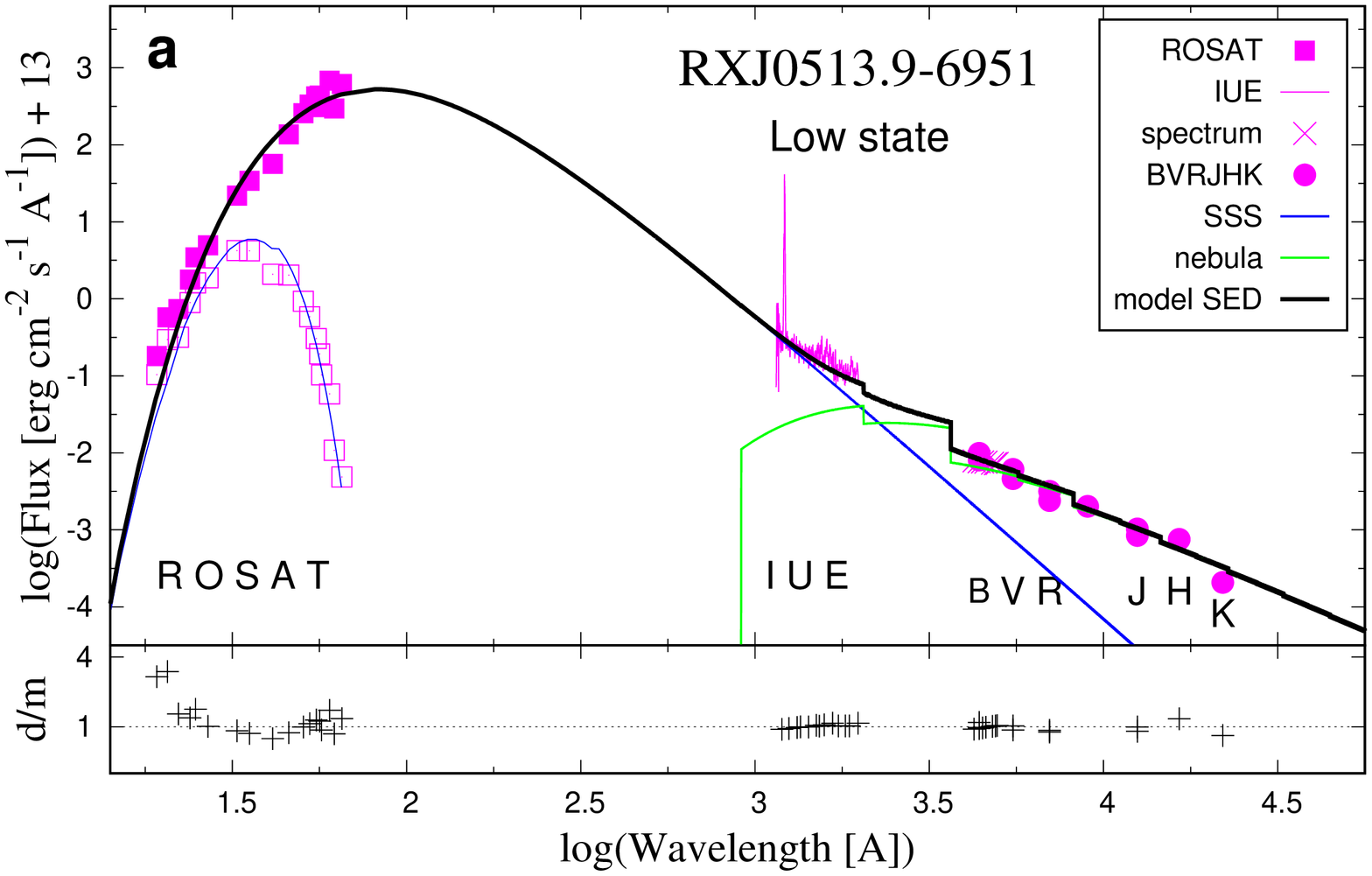}\hspace*{2mm}
           \includegraphics[angle=-90]{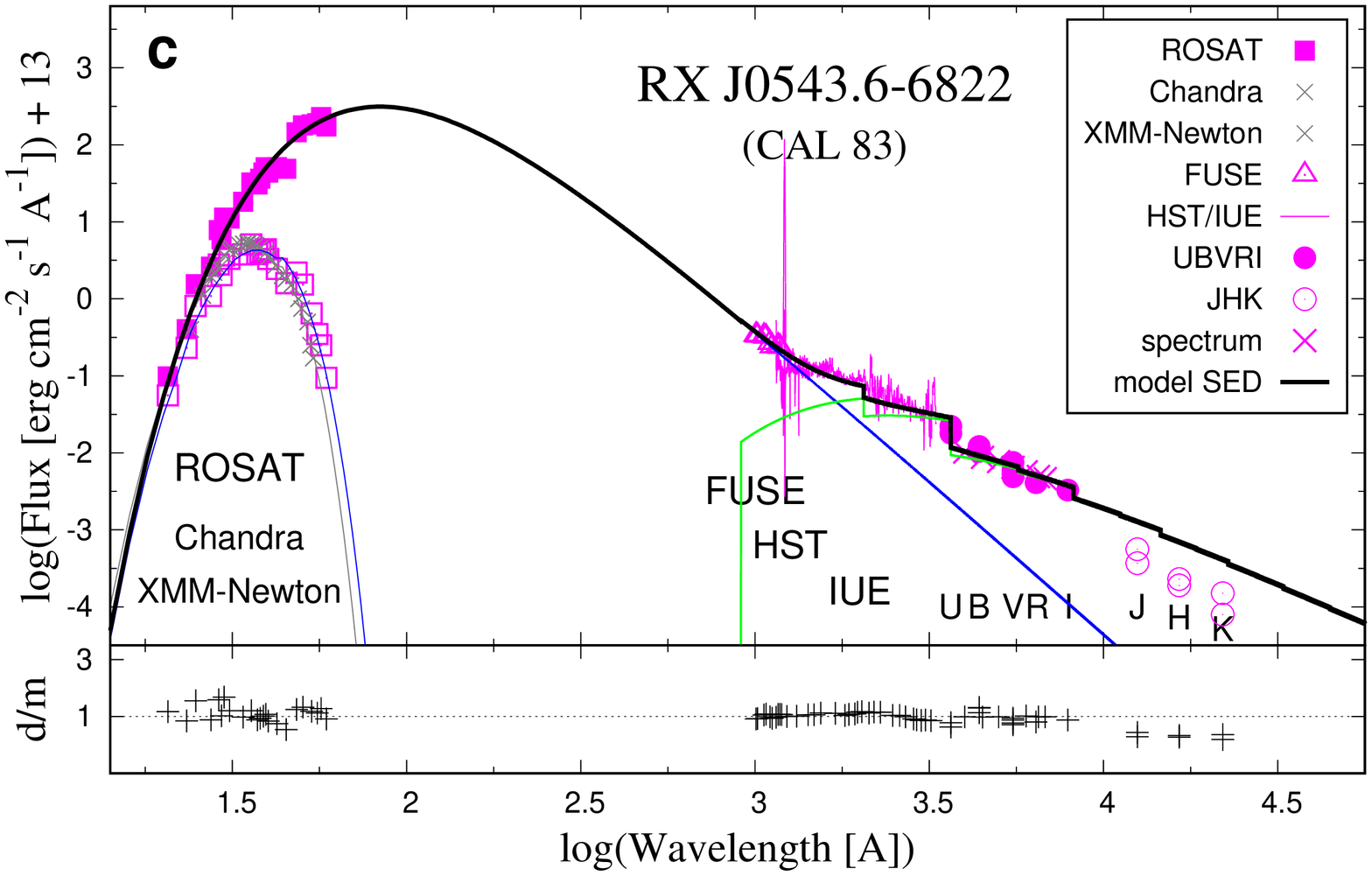}}\vspace*{2mm}
\resizebox{\hsize}{!}
          {\includegraphics[angle=-90]{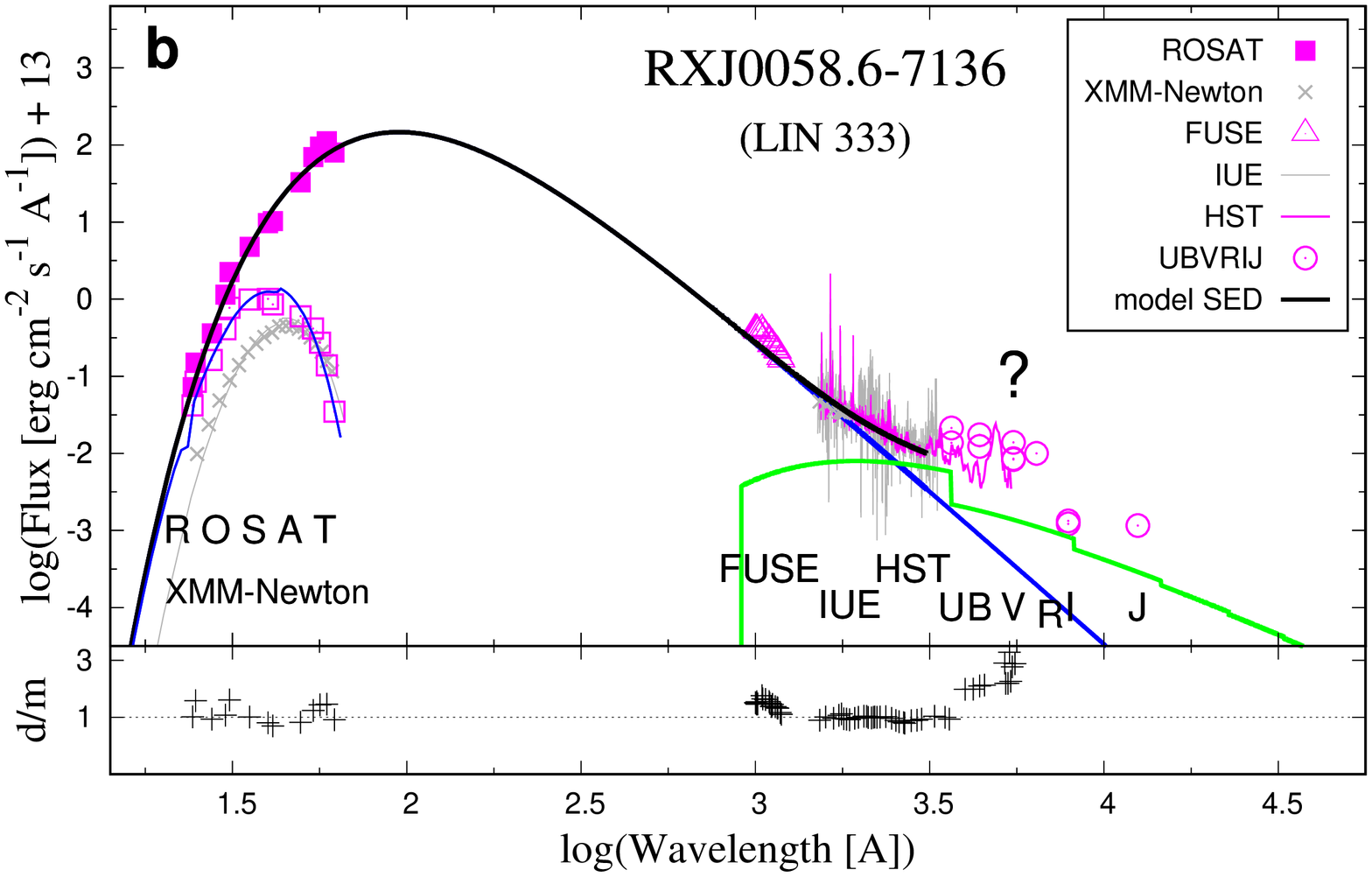}\hspace*{2mm}
           \includegraphics[angle=-90]{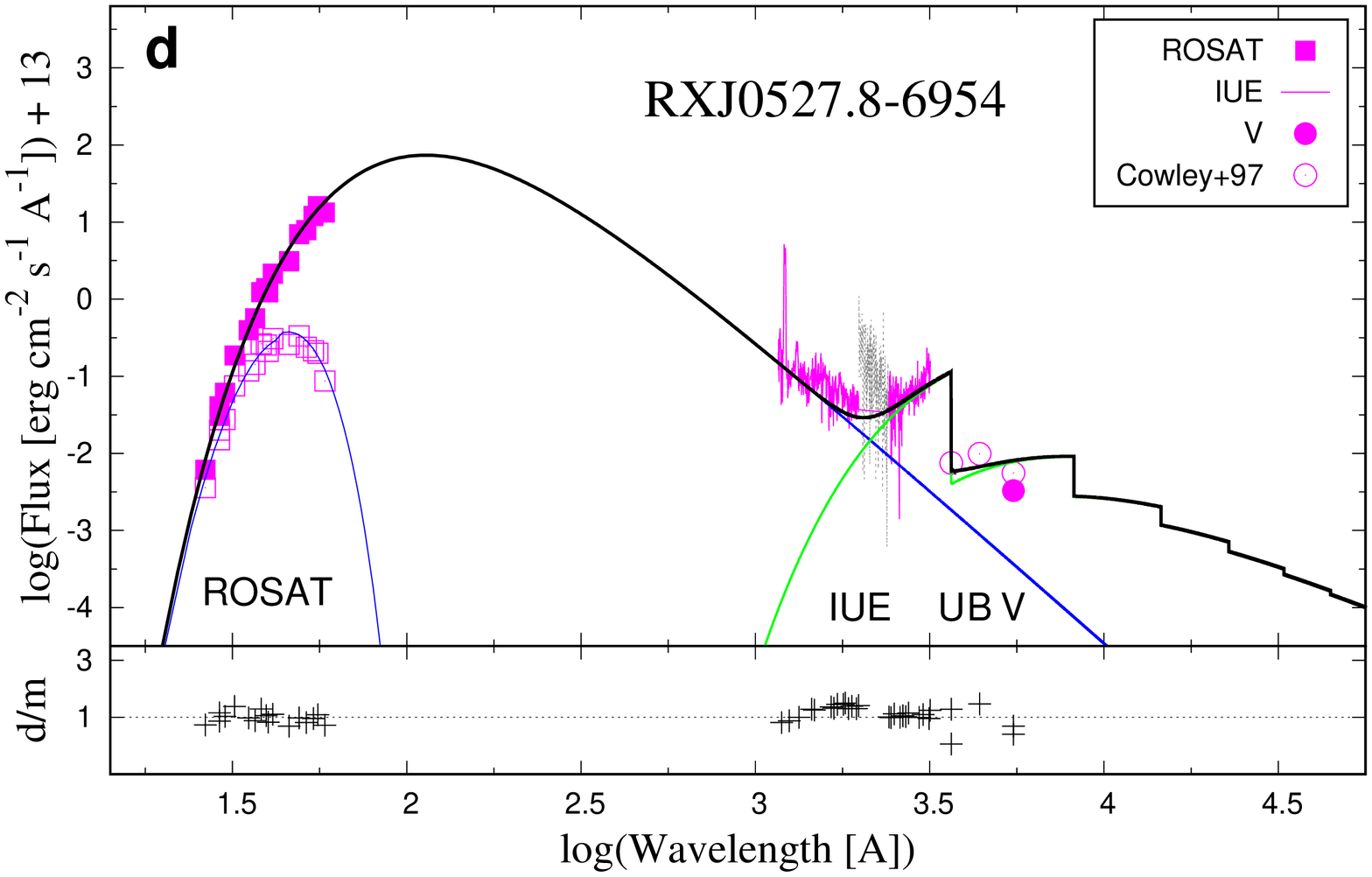}}
 \end{center}
\caption{
A comparison of the measured (in magenta, Sect.~\ref{s:obs}, 
Table~\ref{tab:log}) and model SEDs (heavy black line) of our 
targets with corresponding data-to-model ratios (d/m). 
Filled/open squares are the unabsorbed/observed X-ray fluxes. 
The blue and green lines denote components of radiation from 
the SSS and the nebula, respectively (see Sect.~\ref{s:analysis}, 
Eq.~(\ref{eq:sed2})). 
{\bf a} -- RX\,J0513.9-6951. 
{\bf b} -- RX\,J0058.6-7135 (LIN~333): Two models fitting the 
\textsl{ROSAT}---UV and \textsl{XMM-Newton}---UV fluxes were 
performed. $U,B,V,R$ fluxes and the optical part of the \textsl{HST} 
spectrum suggest the presence of an additional source of radiation 
(marked by {\bf ?}, see Appendix~\ref{s:appC}). 
{\bf c} -- RX\,J0543.6-6822 (CAL~83): Models performed for 
the \textsl{ROSAT} and \textsl{Chandra/XMM-Newton} data 
(each complemented with the same UV and optical fluxes), are 
very similar (Table~\ref{tab:par}). The $J,H,K$ fluxes were 
not used for the modeling. 
{\bf d} -- RX\,J0527.8-6954: Dotted line is the underexposed 
part of the LWP23826 spectrum. 
}
\label{fig:seds}
\end{figure*}

Ultraviolet spectra of RX\,J0527.8-6954 were carried out 
with the \textsl{IUE} satellite on September 1, 1992 
(SWP45499, LWP23826). The LWP23826 spectrum was underexposed, 
therefore its part from the short-wavelength edge to 
$\sim 240$\,nm was not used in the modeling 
(dotted line in Fig.~\ref{fig:seds}d). 

The optical brightness of RX\,J0527.8-6954 was measured by 
\cite{1997PASP..109...21C} in $UBV$ passbands at the Cerro Tololo 
Inter-American Observatory during December 1992, December 1993 
and November 1994. They determined $V = 17.3$, $B = 17.4$ 
and $U = 17.2$.
On September 2004, \cite{2010A&A...517L...5O} observed RX\,J0527.8-6954 
using the Gemini South Telescope with the Gemini Multi Object 
Spectrograph and distinguished the original star into 
two components. They determined the flux ratio in the V band 
between the components to 1.43, and corrected the $V$ magnitude 
of the RX\,J0527.8-695 optical counterpart to 17.9. 
%
%
%
%
\section{Analysis and results}
\label{s:analysis}
\subsection{Modeling the SED}
\label{ss:sed}
The primary aim of this paper is to reconstruct the SED of bright 
SSSs in LMC and SMC from supersoft X-rays to the optical or NIR. 
In this way to determine physical parameters of the main components 
of radiation in the observed continuum. To achieve this aim we use 
the method of multiwavelength modeling the composite continuum of 
SSSs described by \cite{2015NewA...36..116S}. Here we introduce 
its basic assumptions and relationships. 

In our SED modeling the observed continuum, $F(\lambda)$, is 
assumed to consist of a stellar component of radiation, 
$F_{\rm SSS}(\lambda)$, from the WD pseudo-photosphere, and 
nebular continuum, $F_{\rm N}(\lambda)$, from the ionized 
out-flowing gas (see Appendix~\ref{s:appA} and 
Sect.~\ref{ss:dotM}), i.e., 
%
\begin{equation}
  F(\lambda) =  F_{\rm SSS}(\lambda) + F_{\rm N}(\lambda). 
\label{eq:sed1}
\end{equation}
In the model we compare the stellar continuum with the blackbody 
radiation at a temperature $T_{\rm BB}$, while the nebular 
continuum is approximated by contributions from f--b and 
f--f transitions in hydrogen plasma radiating at the electron 
temperature $T_{\rm e}$. 
In cases, where observations suggest a discontinuity in the 
continuum around $\lambda 2050$\,\AA, contributions from 
doubly ionized helium are also included, and the He$^{++}$ 
abundance is set to 0.1. 
According to Eqs.~(6) and (8) of \cite{2015NewA...36..116S}, 
Eq.~(\ref{eq:sed1}) can be expressed in the form,
%
\begin{equation}
  F(\lambda) =   
    \theta_{\rm SSS}^2 \pi B_{\lambda}(T_{\rm BB})
    e^{-\sigma_{\rm X}(\lambda)\,N_{\rm H}} + 
    k_{\rm N}\times\varepsilon_{\lambda}({\rm H,He^{+}},T_{\rm e}), 
\label{eq:sed2}
\end{equation}
where 
$\theta_{\rm SSS} = R_{\rm SSS}^{\rm eff}/d$ is the angular radius 
of the SSS, given by its effective radius (i.e. the radius of 
a sphere with the same luminosity) and the distance $d$. 
Attenuation of the light in the X-ray domain is given by the 
total cross-section for photoelectric absorption per hydrogen 
atom, $\sigma_{\rm X}(\lambda)$ \citep[][]{1974ApJ...187..497C}, and 
the total column density of neutral atoms of hydrogen, $N_{\rm H}$ 
(i.e., given by both the interstellar and circumstellar matter 
component). 
The second term at the right is the nebular continuum 
expressed by its volume emission coefficient 
$\varepsilon_{\lambda}({\rm H,He^{+}},T_{\rm e})$ scaled with 
the factor $k_{\rm N}= EM/4\pi d^2$, where 
$EM=\int_{V}n_{\rm e}n^{+}{\rm d}V$ is the emission measure 
of the nebula given by the electron and a ion concentration, 
$n_{\rm e}$ and $n^{+}$, in the volume $V$ of the ionized 
element under consideration. 
The assumption that the nebular radiation in the continuum can 
be characterized by a single $T_{\rm e}$ is supported by the 
velocity distribution of free electrons in nebulae that tends 
to be Maxwellian, because the electrostatic encounters of free 
electrons are much more likely than any other inelastic 
scatterings \citep[see][]{1947ApJ...105..131B}. 
Therefore, using a single-temperature nebula in the SED modeling 
expresses the measured UV--NIR fluxes quite well 
\citep[see the SED models in][and Fig.~\ref{fig:seds} here]{
2005A&A...440..995S,2015NewA...36..116S,2019ApJ...878...28S}. 
In the modeling procedure we simultaneously fit absorbed 
X-ray continuum fluxes and dereddened UV to NIR fluxes 
searching for parameters $\theta_{\rm SSS}$, $T_{\rm BB}$, 
$N_{\rm H}$, $T_{\rm e}$ and $k_{\rm N}$ 
\citep[see][in detail]{2015NewA...36..116S}\footnote{The 
corresponding software and application example are available 
at \url{https://doi.org/10.5281/zenodo.6985175}}. 
%
The luminosity of the SSS is calculated as 
%
\begin{equation}
  L_{\rm SSS} = 4\pi d^2 \theta_{\rm SSS}^2 \sigma T_{\rm BB}^4 .
\label{eq:lsss}
\end{equation} 
Physical parameters corresponding to our SED models are 
collected in Table~\ref{tab:par}. 
%

\subsection{$N_{\rm H}$ from Rayleigh scattering}
\label{ss:ray}
The high-resolution \textsl{HST/GHRS} spectra ($\lambda$115--144\,nm) 
show a pronounced attenuation of the continuum around the Ly-$\alpha$ 
line that is caused by Rayleigh scattering on atomic hydrogen. 
This effect can be used to independently determine $N_{\rm H}$ in 
the direction of the SSS 
\citep[see Sect.~2.3.1 of][]{2015NewA...36..116S}. 
For this purpose we compare the function, 
\begin{equation}
 F(\lambda) =   
   \theta_{\rm SSS}^2 \pi B_{\lambda}(T_{\rm BB})\,
    e^{-\sigma_{\rm Ray}(\lambda)\,N_{\rm H}} + 
    k_{\rm N}\times\varepsilon_{\lambda}({\rm H},T_{\rm e}),
\label{eq:nhray}
\end{equation}
to the observed spectrum. The Rayleigh cross-section for scattering 
by hydrogen in its ground state, $\sigma_{\rm Ray}(\lambda)$, is 
approximated by Eq.~(5) of \cite{1989A&A...211L..27N}. Other 
parameters have the same meaning as in Eq.~(\ref{eq:sed2}), and 
are fixed according to the solution of the global SED. 
%

\subsection{Multiwavelength model SED of RX\,J0513.9-6951}
\label{ss:0513sed}
\begin{figure*}[t]
 \begin{center}
\resizebox{16cm}{!}{\includegraphics[angle=-90]{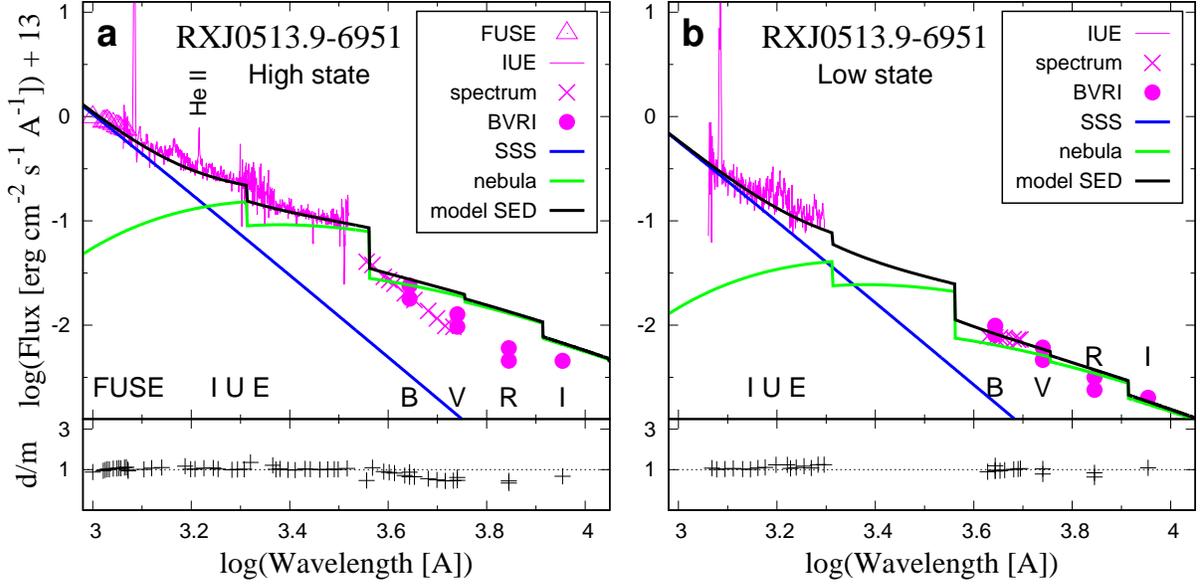}}
 \end{center}
\caption{
The UV--optical SED during the optical-high ({\bf a}) and low 
({\bf b}) state of RX\,J0513.9-6951. This demonstrates that 
the pronounced optical variability with $\Delta V \approx 1$\,mag 
{\bf can be} caused by the variable nebular emission 
(Sect.~\ref{ss:0513sed}). 
}
\label{fig:rx0513sed2}
\end{figure*}

Owing to the strict soft X-ray/optical flux anticorrelation, 
it was possible to perform the X-ray---NIR model SED only 
for the low state of RX\,J0513.9-6951 (Fig.~\ref{fig:seds}a), 
while during the high state, only the far-UV--NIR observations 
were available to model the corresponding SED 
(Fig.~\ref{fig:rx0513sed2}). 
The best fitting X-ray---NIR model SED parameters, 
$N_{\rm H} = (1.2\pm 0.2)\times 10^{21}$\cmd, 
$\theta_{\rm SSS} = 8.8\times 10^{-14}$ 
and 
$T_{\rm BB} = (350000\pm 10000)$\,K, 
correspond to a very high luminosity of the SSS source, 
$L_{\rm SSS} = (1.9\pm 0.7)\times 10^{39}$\es, 
with the effective radius, 
$R_{\rm SSS}^{\rm eff} = (0.19\pm 0.02)$\ro. 
The model SED shows also a significant contribution from 
the thermal nebula with the emission measure, 
$EM = (2.3\pm 0.2)\times 10^{60}$\cmt\ that dominates 
the spectrum for $\lambda \gtrsim 2000$\,\AA. 

During the optical-high state, the nebular emission increased 
by a factor of $\sim$3.5 (Table~\ref{tab:par}), and the WD's 
effective radius inflated to $(0.26\pm 0.03)$\ro\ for the same 
$T_{\rm BB}$ as during the optical-low state. 
The former suggests that the variable nebular emission is 
responsible for the optical variability with 
$\Delta V \approx 1$\,mag, while the latter reflects the 
increase of the far-UV fluxes of the stellar component of 
radiation (the blue lines in Fig.~\ref{fig:rx0513sed2}, 
see Sect.~\ref{ss:anti} in detail). 

\begin{figure}
\begin{center}
\resizebox{\hsize}{!}{\includegraphics[angle=-90]{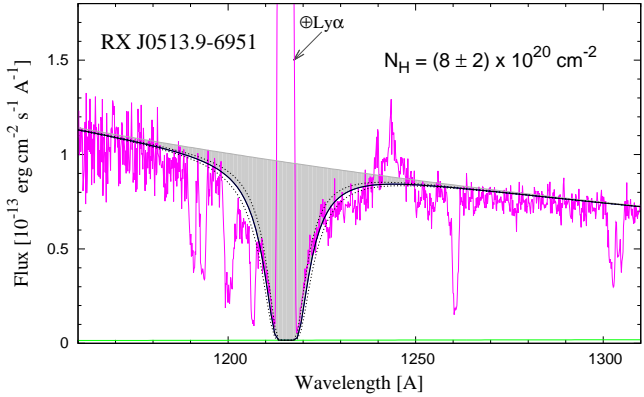}}
\end{center}
\caption{
The \textsl{HST/GHRS} spectrum of RX\,J0513.9-6951 dereddened 
with $E_{\rm B-V} = 0.11$ (in magenta). 
A strong depression of the continuum around the Ly-$\alpha$ 
line (grey area) is caused by Rayleigh scattering of the SSS 
radiation on hydrogen atoms with a column density of 
$(8\pm 2)\times 10^{20}$\cmd\ (black and dotted lines). 
Meaning of blue and green lines as in Fig.~\ref{fig:seds}. 
Geocoronal Ly-$\alpha$ line is denoted by $\oplus$. 
}
\label{fig:rx0513ray}
\end{figure}

Finally, fitting the far-UV \textsl{HST/GHRS} spectrum by 
Eq.~(\ref{eq:nhray}) we found that the hollow around Ly-$\alpha$ 
corresponds to $N_{\rm H} = (8\pm 2)\times 10^{20}$\cmd. 
A comparison of the model SED and the \textsl{HST} spectrum
constrains the reddening $E_{\rm B-V} = 0.11 \pm 0.02$. 
The spectrum and model (\ref{eq:nhray}) are shown in 
Fig.~\ref{fig:rx0513ray}. 

\subsection{Multiwavelength model SED of RX\,J0058.6-7135}
\label{ss:0058sed}
Observations of LIN~333 span the wavelength range from supersoft 
X-rays to the optical. Fluxes in the near-UV and $UBV$ bands 
($\sim 1\times 10^{-15}$\ecsa, see Fig.~\ref{fig:seds}b), 
suggest a contribution from the nebula and, possibly, from 
a companion to the SSS in a binary 
(see Appendix~\ref{s:appC}). 
In modeling the supersoft X-ray---optical SED we fitted 
13 flux-points between 24 and 62\,\AA\ and 31 flux-points 
between 1500 and 3600\,\AA\ by Eq.~(\ref{eq:sed2}). 
Observations with \textsl{FUSE} were not used directly in 
the SED-fitting procedure, because their continuum level is 
a factor of 1.1--1.5 above that given by the \textsl{IUE} 
and \textsl{HST} spectra. 
Nevertheless, the steep slope of their dereddened fluxes 
follows the short-wavelength ends of the \textsl{IUE} and 
\textsl{HST} spectra, which documents dominant contribution 
from the SSS to the far-UV (compared are selected fluxes from 
the LiF2A channel spectrum (1087 -- 1181\,\AA)). 

The modeling solution with the \textsl{ROSAT} data has the 
reduced $\chi^{2} = 2.6$, and corresponds to the fitting 
parameters of the SSS, 
$\theta_{\rm SSS} = (6.6\pm 0.7)\times 10^{-14}$, 
$T_{\rm BB} = 305\,000 \pm 10\,000$\,K and
$N_{\rm H} = (9.0 \pm 0.2) \times 10^{20}$\cmd. 
These parameters imply the effective radius, 
$R_{\rm SSS}^{\rm eff} = 0.17\pm 0.02 (d/60\,\kpc)$\ro, 
and the luminosity, 
$L_{\rm SSS} = 9.1\pm 3.0\times 10^{38}(d/60\,\kpc)^2$\es,
which translates into the unabsorbed X-ray luminosity 
$L_{\rm X}(0.2-1.0\,{\rm keV}) = 
      (4.6 \pm 1.6)\times 10^{37} (d/60\,\kpc)^2$\es.
Using the average fluxes of the three \textsl{XMM-Newton} 
spectra \citep[see Table~\ref{tab:log}, 
Sect.~\ref{ss:0058obs},][]{2010A&A...519A..42M} 
and the same UV--optical 
fluxes as above, yield the following parameters of 
the SSS: 
 $N_{\rm H} = (7.1 \pm 0.2) \times 10^{20}$\cmd, 
 $T_{\rm BB} = 267\,000 \pm 7\,000$\,K, 
 $R_{\rm SSS}^{\rm eff} = 0.18\pm 0.02 (d/60\,\kpc)$\ro\ 
and 
 $L_{\rm SSS} = 5.7\pm 1.8\times 10^{38}(d/60\,\kpc)^2$\es. 
%

The nebular component of radiation dominates the spectrum 
from around 2500\,\AA\ to longer wavelengths 
(Fig.~\ref{fig:seds}b). Its amount is scaled with 
$k_{\rm N} = 3.8\pm 0.4\times 10^{12}\,{\rm cm^{-5}}$,
which corresponds to 
$EM = (1.6\pm 0.2)\times 10^{60}(d/60\,\kpc)^2$\cmt.
The nebula radiates at a high $T_{\rm e} = 37\,000\pm 5000$\,K.
Both, $EM$ and $T_{\rm e}$ are in a good agreement with those 
derived by \cite{1987ApJ...320..159A} from emission lines: 
$EM \sim 1.8\times 10^{60}(d/63.1\,\kpc)^2$\cmt, 
$T_{\rm e} \sim 25\,300 - 31\,600$\,K (see their Tables~5 
and 7, Appendix~\ref{s:appA} here). 

Finally, it is of interest to note that the long time-scale 
of observations of LIN~333 (1984 -- 2009, Table~\ref{tab:log}) 
with comparable X-ray and UV fluxes demonstrates a stability 
of this bright SSS \citep[][]{2010A&A...519A..42M}. 
%
%
%
%
%
%
%
%
\begin{figure}
 \begin{center}
\resizebox{\hsize}{!}{\includegraphics[angle=-90]{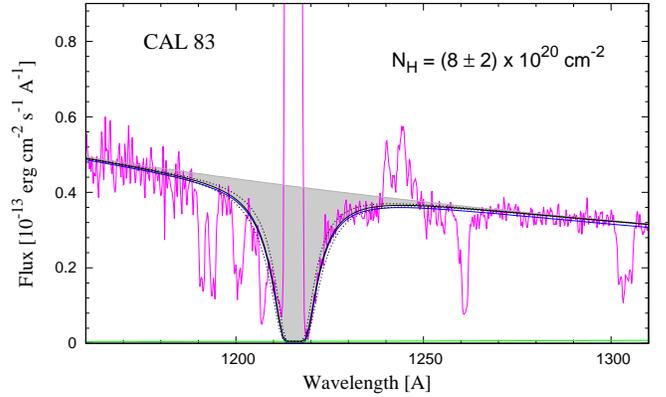}}
 \end{center}
\caption{
As in Fig.~\ref{fig:rx0513ray}, but for CAL~83. The spectrum 
is dereddened with $E_{\rm B-V} = 0.11$.
}
\label{fig:cal83ray}
\end{figure}
%

\subsection{Multiwavelength model SED of CAL~83}
\label{ss:cal83sed}
As in the case of RX\,J0513.9-6951, the high values of the 
near-UV--optical fluxes (a few $\times 10^{-15}$\ecsa), their 
steepness (e.g., $U-B \sim -1$\,mag) and the presence of 
emission lines of the hydrogen Balmer series, \heii\,$\lambda$4686, 
N\I\I\I\,4640\,\AA\ and [O\I\I\I]\,5007\,\AA\ 
in the optical spectrum \citep[e.g.,][]{1985SSRv...40..229P} 
indicate a strong nebular continuum that dominates 
the spectrum for $\lambda > 1\,800$\,\AA\ 
(see Fig.~\ref{fig:seds}c). 

We modeled the SED of CAL~83 between 20.7 and 9\,000\,\AA\ 
by Eq.~(\ref{eq:sed2}). The best model with \textsl{ROSAT} 
data has the reduced $\chi^{2} = 3.8$ (for 61 d.o.f.) and 
determines the variables of the SSS, 
 $\theta_{\rm SSS} = (6.9\pm 0.6)\times 10^{-14}$,
 $T_{\rm BB} = 345\,000 \pm 15\,000$\,K and 
 $N_{\rm H} = (1.03\pm 0.05)\times 10^{21}$\cmd,
which yields 
 $R_{\rm SSS}^{\rm eff} = 0.15\pm 0.02 (d/49\,{\rm kpc})$\ro\
and
 $L_{\rm SSS} = (1.1\pm 0.5)\times 10^{39}
                (d/49\,{\rm kpc})^2$\es. 
Model variables of the nebular component of radiation, 
 $k_{\rm N} = (1.0\pm 0.1)\times 10^{13}$\,cm$^{-5}$ and
 $T_{\rm e} = 30\,000 + 20\,000/-10\,000$\,K 
correspond to 
 $EM = (2.9\pm 0.3)\times 10^{60}(d/49\,{\rm kpc})^2$\cmt\ 
(Table~\ref{tab:par}). 

The model SED with \textsl{Chandra/XMM-Newton} fluxes is very 
similar to that with the \textsl{ROSAT} data 
(see Fig.~\ref{fig:seds}c, Table~\ref{tab:par}). 
The corresponding parameters of the SSS are, 
 $N_{\rm H} = (1.3 \pm 0.05) \times 10^{21}$\cmd, 
 $T_{\rm BB} = 357\,000 \pm 15\,000$\,K, 
 $R_{\rm SSS}^{\rm eff} = 0.16\pm 0.02 (d/49\,\kpc)$\ro\  
and 
 $L_{\rm SSS} = (1.4\pm 1.1)\times 10^{39}(d/49\,\kpc)^2$\es. 
As in the case of LIN~333, the long time-scale of observations 
demonstrate that CAL~83 is a stable SSS during its X-ray-on 
states. 
%

Finally, the attenuation of the continuum around Ly-$\alpha$, 
observed on the \textsl{HST/GHRS} spectrum, corresponds to 
$N_{\rm H} = (8 \pm 2)\times 10^{20}$\cmd\ 
(see Fig.~\ref{fig:cal83ray}). 
\subsection{Multiwavelength model SED of RX\,J0527.8-6954}
\label{ss:0527sed}
The steepness of the SWP45499 spectrum suggests that the SSS 
radiation dominates also a wide range of the far-UV continuum 
(115--190\,nm), while the LWP23826 spectrum has an opposite 
slope, whose profile suggests a dominant contribution from 
a low temperature nebula (see Fig.~\ref{fig:seds}d). 
Available observations allow us to model the continuum 
fluxes between 26.5 and 5\,500\,\AA. 
The best fit to the X-ray and UV data (39 fluxes) has 
the reduced $\chi^{2} = 3.3$ (for 34 d.o.f.) and determines 
the model variables, 
 $\theta_{\rm SSS} = (7.3\pm 0.7)\times 10^{-14}$,
 $T_{\rm BB} = 255\,000 \pm 20\,000$\,K and
 $N_{\rm H} = (7.1\pm 0.2)\times 10^{20}$\cmd, 
which yield 
 $R_{\rm SSS}^{\rm eff} = 0.16\pm 0.02 
                        (d/49\,{\rm kpc})$\ro\
and
 $L_{\rm SSS} = (3.7\pm 0.5)\times 10^{38}
              (d/49\,{\rm kpc})^2$\es.
The nebular component of radiation is determined by 
 $k_{\rm N} = (1.5\pm 0.1)\times 10^{13}$\,cm$^{-5}$ and
 $T_{\rm e} = 9\,000\pm 2\,000$\,K that gives 
 $EM = (4.3\pm 0.3)\times 10^{60}(d/49\,{\rm kpc})^2$\cmt\ 
(Table~\ref{tab:par}). 

\section{Discussion}
\label{s:dis}
\subsection{Amount of absorption to the SSSs}
\label{ss:nh}
Independently of the SED modeling, we estimated the value of 
$N_{\rm H}$ also from Rayleigh scattering of the SSS radiation on 
H atoms that creates a strong depression of the continuum around 
the Ly$\alpha$ line (see Figs.~\ref{fig:rx0513ray} and 
\ref{fig:cal83ray}). 
The larger inaccuracy of this method is given by a weaker 
sensitivity of the Rayleigh scattering process to the amount 
of scatterers in the Ly$\alpha$ wings, and by the complexity 
of the depression profile due to superposition of rich 
absorption-line spectrum to the continuum. 
On the other hand, determination of $N_{\rm H}$ from the SED 
modeling suffer by rather uncertain X-ray fluxes taken from 
already published figures and their extreme sensitivity to 
the amount of absorption within the supersoft X-ray wavelengths. 
Another source of uncertainties in determining values of 
$N_{\rm H}$ is given by using non-simultaneous X-ray and UV 
observations, the temporal variability of the SSS radiation, 
and/or the circumstellar component of $N_{\rm H}$ originating 
in the surrounding nebula (Sect.~\ref{ss:anti}, 
Appendix~\ref{s:appA}). 
Nevertheless, $N_{\rm H}$ values obtained by these two different 
methods are comparable (see Table~\ref{tab:par}, and 
Figs.~\ref{fig:rx0513ray} and \ref{fig:cal83ray} for 
RX\,J0513.9-6951 and CAL~83). 
In the case of LIN~333 and CAL~83 we modeled different X-ray 
data together with the same UV--optical observations 
(Table~\ref{tab:par}, Figs.~\ref{fig:seds}b and \ref{fig:seds}c). 
Despite this shortcoming, the $N_{\rm H}$ values we found for 
LIN~333 are in good agreement with the total column density 
$N_{\rm H}\approx (7\pm 4)\times 10^{20}$\cmd\ determined by 
\cite{1991MNRAS.252P..47W}, which includes contributions in 
the Galaxy, in the SMC, and inside the nebula. 
For RXJ0527.8-6954 we used X-ray fluxes from June 1990 and 
probably fainter far-UV fluxes from September 1992 as 
suggested by the X-ray and optical light curves 
\citep[see][]{1996A&A...312...88G,1997MNRAS.291L..13A}. 
If this is so, the true value of $N_{\rm H}$ should be little 
higher than that given in Table~\ref{tab:par}. 
%
%
\begin{figure*}
 \begin{center}
\resizebox{16cm}{!}{\includegraphics[angle=-90]{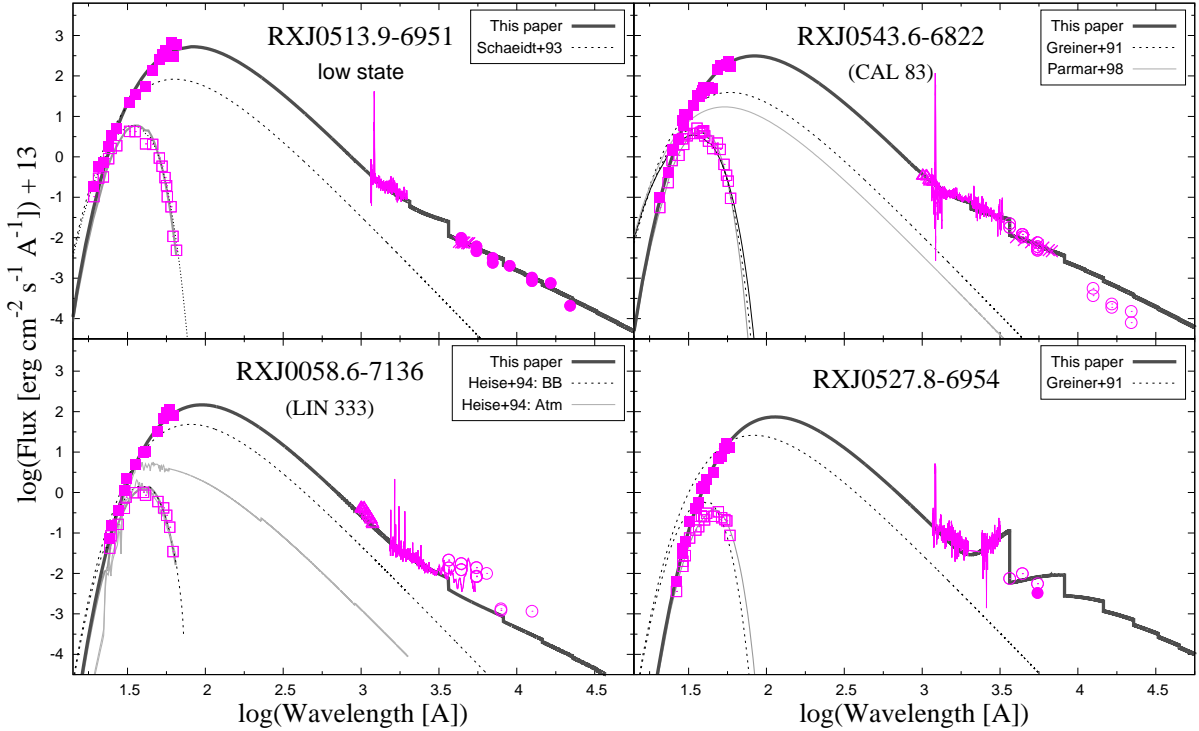}}
 \end{center}
\caption{
Comparison of the multiwavelength SED models of this paper (dark 
thick lines, Table~\ref{tab:par}) and the corresponding X-ray-data 
blackbody models from the literature (see legends, 
Table~\ref{tab:lit}, Sect.~\ref{ss:comp}). 
}
\label{fig:comparison}
\end{figure*}

\subsection{Comparison with previous models}
\label{ss:comp}
Figure~\ref{fig:comparison} compares our multiwavelength models 
and the X-ray-data models from the literature described in 
Sects.~\ref{ss:0513} to \ref{ss:0527}, and summarized in 
Table~\ref{tab:lit}. 
In both cases, the SED models match satisfactorily the absorbed 
X-ray fluxes. However, the unabsorbed models differ, often 
significantly, from our multiwavelength models. Mostly, they are 
a factor of $\sim 10-100$ below the far-UV data. 
Such the discrepancy between the multiwavelength modeling and 
the modeling only the X-ray data can be ascribed to the mutual 
dependence between $N_{\rm H}$, $L_{\rm SSS}$, and $T_{\rm BB}$ 
parameters, as described in Sect.~4.1 of 
\cite{2015NewA...36..116S}. 

RX\,J0513.9-6951: The X-ray blackbody solution of 
\cite{1993A&A...270L...9S} corresponds 
to $L_{\rm SSS} = 2.3\times 10^{38}$\es, which is a factor of 
$\sim 15$ below the far-UV fluxes. The model atmosphere fit 
by \cite{1998A&A...333..163G} yields even lower luminosity of 
$(2.5 - 9)\times 10^{37}$\es\ (see Sect.~\ref{ss:0513}). 

RX\,J0058.6-7135 (LIN~333): Blackbody and atmospheric X-ray SED 
models performed by \cite{1994A&A...288L..45H} are a factor 
of $\sim$9 and $\sim$300 below the \textsl{FUSE} data, 
respectively. 
They were already discussed by \cite{2015NewA...36..116S} as 
an illustrative example of the mutual dependence of the model 
parameters, $N_{\rm H}$, $L_{\rm SSS}$ and $T_{\rm BB}$ 
(see Sect.~4.1 and Fig.~7 there). 
Similar results were obtained also by \cite{2010A&A...519A..42M} 
using the \textsl{XMM-Newton} observations 
(see Table~\ref{tab:lit}). 
The authors questioned their atmospheric model when they found 
that the flux predicted in the visible band is lower by $\sim$30\% 
than the upper limit ($V\sim 21$) obtained from the \textsl{HST} 
imaging of the central star by \cite{2004ApJ...614..716V}. 

RX\,J0543.6-6822 (CAL~83): Blackbody solution at the Eddington 
limit \citep[$N_{\rm H} = 8\times 10^{20}$\cmd, 
     $kT_{\rm BB} = 43$\,eV, 
     $R_{\rm BB} = 1.5\times 10^9$\,cm,][]{1991A&A...246L..17G} 
%
%
is a factor of $\sim 20$ below the far-UV continuum, while 
their softer model 
($N_{\rm H} = 1.7\times 10^{21}$\cmd, $kT_{\rm BB} = 26$\,eV) 
fits the absorbed X-ray data for 
$\theta_{\rm SSS}\sim 4\times 10^{-13}$ 
(i.e., $R_{\rm BB} = 6.0\times 10^{10}$\,cm, 
$L_{\rm SSS} \sim 2\times 10^{40}$\es) 
that is a factor of $\sim 30$ above the far-UV fluxes. 
Independent modeling of the \textsl{BeppoSAX} spectrum of 
CAL~83 by \cite{1998A&A...332..199P} also matches 
well the observed X-ray data, but the unabsorbed model is 
by a factor of $\sim 80$ below the far-UV continuum. 

RX\,J0527.8-6954: Similar results were obtained also for 
this SSS by \cite{1991A&A...246L..17G}. 

\begin{figure}
 \begin{center}
\resizebox{\hsize}{!}{\includegraphics[angle=-90]{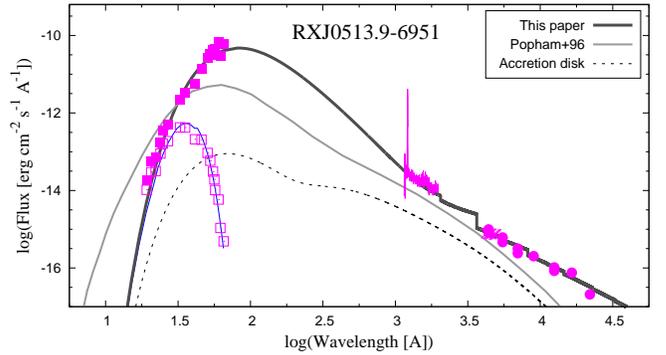}}
 \end{center}
\caption{
Comparison of our SED model for RX\,J0513.9-6951 and that 
including the steady-burning WD at the Eddington limit 
irradiating the flared accretion disk and the donor star 
\citep[gray line, adopted from Fig.~4 of][]{1996LNP...472...65P}. 
The model is aimed to explain the strong UV and optical 
fluxes. Compared is a spectrum of accretion disk with no 
external heating (dashed line, Sect.~\ref{ss:comp}). 
}
\label{fig:compAD}
\end{figure}

Figure~\ref{fig:compAD} compares our SED model for RX\,J0513.9-6951 
and that of \cite{1996LNP...472...65P}, 
who calculated models of irradiated flared accretion disks around 
luminous steady-burning WDs in bright LMC SSSs (CAL~83, CAL~87, and 
RX\,J0513.9-6951) and compared the resulting disk fluxes to optical 
and UV observations. 
To increase the radiation from the disk at these wavelengths, 
the disk has to be flared to reprocess sufficiently large amount 
(more than 1/4) of the WD radiation into the optical and UV 
(see their Fig.~2). 
Figure~\ref{fig:compAD} shows their resulting model (gray line) 
that consists of three components of radiation: 
  (i) From a 1.2\mo\ steady-burning WD 
($\dot M_{\rm acc} = 5\times 10^{-7}$\myr, 
$L_{\rm WD} = 1.5\times 10^{38}$\es, 
$T_{\rm BB} = 5\times 10^{5}$\,K and 
$R_{\rm WD} = 1.84\times 10^{9}$\,cm),
  (ii) from the irradiated disk, whose heated surface increases 
the optical and UV fluxes by a factor of 3--5, while the heated 
star-disk boundary makes much brighter the high-energy spectrum, 
and 
  (iii) from the irradiated fraction of the donor star that 
contributes mainly in the optical spectrum. 
For a comparison, the disk spectrum with no external heating 
corresponding to the above-mentioned parameters (i.e., the 
accretion luminosity of $2.7\times 10^{36}$\es) and the outer 
disk edge of $1.35\times 10^{11}$\,cm 
\citep[][]{1996LNP...472...65P} is shown by 
the dashed line\footnote{For the sake of simplicity, we adopted 
optically thick disk that radiates locally like a blackbody 
\citep[][]{1995cvs..book.....W}.}. 
Note that a more sophisticated calculation performed by 
\cite{1996LNP...472...65P} provides similar profile of the disk 
spectrum, but it is just somewhat shifted to shorter wavelengths 
(see their Figs.~1 and 2). 
The Popham \& Di Stefano's model is calculated for the disk 
inclination of 60$^{\circ}$. If viewing the disk face-on, its 
flux increases twice, and will match better the measured 
UV and optical fluxes. However, it is rather below the NIR 
fluxes and does not reproduce their slope. 
The model does not match the X-ray fluxes, at all. 

Our multiwavelength SED models of the bright SSSs correspond to 
super-Eddington luminosity even for a high-mass WD accretor. 
The SED from the near-UV to longer wavelengths indicates 
a dominant contribution from thermal nebula. Both results 
are fairly startling with respect to the present picture of 
SSSs, whose luminosities can reach in maximum the Eddington 
limit for a WD accretor, and the UV--optical spectrum is 
thought to be dominated by the light from accretion disk 
irradiated by the central steady-burning WD
\citep[][]{1996LNP...472...65P}. 
Below, we discuss these properties in detail. 
%
%

\subsection{Problem of the high luminosity}
\label{ss:lum}
\subsubsection{Accreting white dwarf ?}
\label{sss:wd}
According to \cite{1992A&A...262...97V}, 
the radiation of SSSs is generated by a stable nuclear burning 
of hydrogen accreted onto massive WDs. This idea is based on 
a theoretical conclusion that there is a small range of accretion 
rates, $\dot M_{\rm acc}$, at which the hydrogen-rich material is 
quasi-steady burning on the WD's surface as it is accreting 
\citep[][]{1978ApJ...222..604P,1982ApJ...257..767F}. 
The material is transported onto the accretor trough an accretion 
disk. Then, the maximum luminosity of a SSS is given by that 
generated by the nuclear burning on the WD surface and the 
gravitational potential energy of the accretor, converted into 
the radiation by the disk and its boundary layer, i.e., 
\begin{equation}
 L_{\rm SSS} = \eta X \dot M_{\rm acc} + 
               \frac{G M_{\rm WD}\dot M_{\rm acc}}{R_{\rm WD}}, 
\label{eq:ltot}
\end{equation}
where 
$\eta\sim 0.007\times c^2 \sim 6.3 \times 10^{18}$\,erg\,g$^{-1}$ 
is the energy production of 1 gram of hydrogen due to the nuclear 
fusion of 4 protons, and $X\equiv0.7$ is the hydrogen mass fraction 
in the accreted matter. $M_{\rm WD}$ and $R_{\rm WD}$ are the WD 
mass and radius. 
However, even the extreme parameters of the WD accretor, 
$M_{\rm WD} \sim 1.4$\mo, $R_{\rm WD} \sim 0.003$\ro\ 
\citep[][]{2000A&A...353..970P} and 
the mass transferred through the disk at a high rate of 
a few times $10^{-7}$\myr\ that sustains the stable H-burning 
\citep[see e.g., Fig.~2 of][]{2007ApJ...660.1444S} generates 
the total luminosity of $\gtrsim 10^{38}$\es, which 
is one order of magnitude below that suggested by our 
multiwavelength models SED (Table~\ref{tab:par}). 
The observed luminosity would require a high accretion rate 
of $\dot M_{\rm acc} \sim 10\,\dot M_{\rm Edd}$, where 
$\dot M_{\rm Edd}$ is the Eddington accretion rate. 
This result is not consistent with the model of steady burning 
WD for the brightest SSSs in LMC and SMC, unless the radiation 
of the source is not isotropic. Here, the emission from steady 
nuclear burning onto a WD should be beamed with a factor, 
$L/L_{\rm sph}\lesssim 0.1$, where $L_{\rm sph}$ is the inferred 
isotropic luminosity and $L$ is the true source luminosity. 
%

\subsubsection{Accreting neutron star ?}
\label{sss:ns}
A neutron star (NS) as the accretor in the brightest SSSs provides 
a somewhat better possibility to explain their high luminosities, 
because of a larger efficiency in converting gravitational 
energy into the heat and radiation. The efficiency of radiative 
emission in units of the total mass energy accreted on 
a compact object is 
\begin{equation}
 \xi = \frac{L_{\rm SSS}}{\dot M_{\rm acc}\,c^2}
     = \frac{G M_{\rm acc}}{R_{\rm acc}\, c^2}. 
\end{equation}
For a 1.4\mo\ NS with a radius of 10--15\,km, 
$\xi \sim 0.2-0.14$, while for a WD (1\mo, 7000\,km), 
$\xi \sim 2\times 10^{-4}$. 
Considering that the efficiency of turning mass energy into 
radiation via the nuclear fusion is of $\sim 7\times 10^{-3}$ 
(Eq.~(\ref{eq:ltot})), the nuclear fusion reactions can be 
a vital source of energy for accreting WDs, but are negligible 
in converting the mass energy accreted by a NS. 

To generate the luminosity of $\sim 10^{39}$\es, the above 
mentioned parameters of a NS require the accretion rate 
$\dot M_{\rm acc} = 0.85-1.3\times 10^{-7}$\myr. 
These values are more realistic than those required by the WD 
accretor. Such the material can be supplied by mass transfer 
from a close Roche-lobe filling companion. However, these 
accretion rates are still above the critical values even 
for a high mass NS. 

The presence of a NS as the accretor in the luminous SSSs 
could be evidenced by e.g., detection of strictly periodic 
short-term light variation on the time-scale of milliseconds 
to seconds, or by direct measuring of the size of the X-ray 
emission region from eclipse analysis 
\citep[][]{2017PASP..129f2001M}. 
However, no such observations indicating these properties of 
SSSs have been reported yet. 
In the cases, where a NS probably accretes from the wind of its 
companion, we measure a significantly harder X-ray spectrum 
dominating 1--10\,keV 
\citep[][]{2007A&A...470..331M,2007A&A...464..277M} 
than it is generated by our brightest SSSs. 
Therefore, the presence of accreting NSs in the brightest LMC 
and SMC SSSs seems unlikely. 

\subsubsection{Accreting black hole ?}
\label{sss:bh}
The high X-ray luminosities $L_{\rm X} \gtrsim 10^{39}$\es\ 
can be interpreted by a sub-Eddington accretion onto 
intermediate-mass black hole \citep[BH,][]{2006MNRAS.372..630F}. 
For example, \cite{2003ApJ...590L..13K} applied this 
possibility to the luminous SSS XMMU\,J005510.7--373855 in the 
spiral galaxy NGC\,300 ($L_{\rm bol} = 0.2-2\times 10^{39}$\es), 
and found that accretion disk around a $\approx$2800\mo\ BH 
radiating at $kT = 60$\,eV is required to generate the luminosity 
of $10^{39}$\es. 
However, such the massive BHs, that are usually connected with 
the star-forming regions \citep[][]{2004ApJ...601L.143I}, are 
unlikely to be present in LMC and SMC. 

On the other hand, \cite{2008MNRAS.385L.113K} considered that the 
high luminosity of ultraluminous X-ray sources (ULXs) is generated 
by a stellar-mass BH accreting material from a close companion star 
at super-Eddington rates ($\dot M_{\rm acc} > \dot M_{\rm Edd}$) 
with some degree of geometric beaming. 
According to \cite{2009MNRAS.397.1836G}, 
a relatively cool SSS photosphere for the super-Eddington 
accretion flows could be created by a massive wind that 
completely envelopes the inner-disc regions. They also 
concluded that all spectral features in their sample of ULXs 
can be explained by stellar mass BHs at high accretion rates. 

However, there are two main observational arguments contradicting 
the presence of BHs in the brightest SSSs in LMC and SMC. 
\begin{itemize}
\item 
First, it is the profile of the X-ray spectrum that is considerably 
harder than that produced by our brightest SSSs 
\citep[][]{2010ApJ...716..181J,2016MNRAS.456.1837S}. 
In addition, the spectral shape of ULXs can show either a single 
peak at 2-3\,keV or two peaks \citep[one below 1\,keV and one 
above 3\,keV, see][]{2009MNRAS.397.1836G}, or otherwise 
structured shape of the spectrum 
\citep[][]{2014Natur.514..198M,2016MNRAS.456.1859U} -- all totally 
different from those, here studied, brightest SSSs. 
\item 
Second, if the jets are observed, their velocities are expected 
to be comparable to the escape velocity of the central compact 
object \citep[][]{1997ASPC..121..845L}. 
Accordingly, jets observed from objects containing a BH accretor 
are accelerated to relativistic speed 
\citep[][]{1999ARA&A..37..409M,
           2006MNRAS.369..603F,
           2010PASJ...62L..43T,
           2015Natur.528..108L}, 
while those launched by a WD accretor reach speeds of the order 
of thousands of \kms\ only 
\citep[][]{1996ApJ...470.1065S,
           1998ApJ...506..880B,
           2009ApJ...690.1222S}. 
\end{itemize}
Therefore, considering the observed X-ray SEDs of our targets, 
and taking into account that jets from RX\,J0513.9-6951 are 
ejected at $\pm$3800\kms\ \citep[][]{1996ApJ...470.1065S}, 
there is no other option than that a WD is present also in 
the most luminous SSSs in LMC and SMC. 

\subsubsection{Luminous post-nova supersoft X-ray sources ?}
\label{sss:postnova}
Due to the above reasons, we assume that the accretor in 
the brightest SSSs, we investigate in this paper, is a WD. 
A possibility how to keep a SSS at the super-Eddington state for 
a long time is to fuel the burning WD after the classical nova 
eruption, when it is in the post-nova SSS state, and is burning 
at/above the nuclear Eddington limit, i.e., producing the 
super-Eddington luminosity. 
Below we support this idea by observations of similar cases, and 
theoretical considerations. 

{\sf Examples from observations.} 
A transient post-nova SSS phase occurs after the optical outburst 
and is of varying duration. The shortest one is measured for 
the 1\,yr recurrent nova M31N 2008-12a in the Andromeda galaxy 
that lasts for about two weeks \citep[][]{2018ApJ...857...68H}, 
while in the cases of GQ~Mus and LMC~1995, the SSS phase was 
detected for $\sim$10\,yr 
\citep[][]{1993Natur.361..331O,1995ApJ...438L..95S} 
and 6--8\,yr \citep[][]{2003ApJ...594..435O}, respectively. 
Long-lasting super-Eddington luminosity was also documented for 
some extraordinary novae, e.g., FH~Ser 
\citep[][]{1987A&A...179..164F}, 
LMC~1988 $\#$1 \citep[][]{1998MNRAS.300..931S}, 
LMC~1991 \citep[][]{2001MNRAS.320..103S}, 
SMCN~2016-10a \citep[][]{2018MNRAS.474.2679A}, 
and nova V339~Del \citep[][]{2019ApJ...878...28S}. 
The unusually long phase of hydrogen burning was indicated for 
the nova V723~Cas (1995) by detecting it as a bright SSS more 
than 12 years after the outburst, with possible super-Eddington 
luminosity 
\citep[depending on the spectral model, see][]{2008AJ....135.1328N}. 
The transient SSS in the nearby galaxy NGC 300, denoted as SSS$_1$, 
has the bolometric luminosity of $\approx10^{39}$\es\
\citep[][]{2003ApJ...590L..13K}. 
The source was found in outburst in 1992, 2000, 2008 and 2016 
suggesting a possible recurrence period of about 8 years, and 
thus could be associated with a recurrent nova 
\citep[][]{2019MNRAS.490.4804C}. 
Recently, \cite{2020MNRAS.499.2007V} 
discovered a 30-yr long-lived post-nova SSS in LMC undergoing 
residual surface nuclear burning. 
A large number of post-nova SSSs (250--600) was identified 
in M31 \citep[][]{2016MNRAS.455..668S}. The authors found that 
depending on the WD mass distribution in novae, their unabsorbed 
X-ray luminosity distribution shows significant steepening around 
$L_{\rm X}{\rm (0.2-1.0\,keV)}\approx 10^{38}$\es\ with a maximum 
at $\approx2\times10^{38}$\es. According to our multiwavelength 
SED modeling, 
$L_{\rm SSS}/L_{\rm X}{\rm (0.2-1.0\,keV)}\gtrapprox 10$ 
(see Table~\ref{tab:par}), which would correspond to the 
bolometric luminosities of the brightest SSSs in M31 of 
$\approx10^{39}$\es. Note, however, that this ratio is a strong 
function of the temperature: 
$L_{\rm SSS}/L_{\rm X}{\rm (0.2-1.0\,keV)}\sim 21$ 
or $\sim$2.4 only for 300\,000 or 600\,000\,K, respectively. 
Finally, long-term accretion at/above the nuclear Eddington limit 
is also indicated for symbiotic X-ray binaries RXJ0059.1-7505 
(LIN~358) in SMC 
\citep[][]{2015NewA...36..116S,
           2021MNRAS.500.3763K,
           2021ApJ...918...19W} 
and Draco~C1 in the Draco dwarf spheroidal galaxy
\citep[][]{2018MNRAS.473..440S,2020ApJ...900L..43L}, where 
the donor star is an evolved red giant. 

{\sf Theoretical considerations.} 
According to \cite{1988ApJ...325..828K} 
the irradiation of the donor star after the nova outburst can 
cause the red dwarf to expand and induce a mass transfer rate 
enhanced by two orders of magnitude. 
If the donor is a red giant (symbiotic binaries), the high 
wind-mass-transfer efficiency is achieved by the wind focusing 
to the orbital plane 
\citep[][]{2015A&A...573A...8S,
           2016A&A...588A..83S,
           2021A&A...646A.116S}. 
In this case, the enhanced accretion can be sustained also 
through the disk warping that leads to ejection of transient 
jets \citep[][]{2018ApJ...858..120S}. 
If the accretion resumes immediately at the beginning of 
the post-nova SSS phase, 
the SSS lifetime can be substantially lengthened
\citep[][]{2017ApJ...838..153K}. Illustration of this effect is in 
Fig.~15 of \cite{2018ApJ...857...68H}. 
At a high accretion rates ($\gtrsim 5\times 10^{-7}$\myr), so 
that a new outburst can occur on the human lifetime, the SSS 
phase can be interrupted for a relatively short time by the 
optical outburst being followed by the SSS phase at very high 
bolometric luminosity. 
Such the type of variability has been observed for the SSS SSS$_1$ 
in NGC 300 (see above), and here for RX\,J0527.8-6954, where 
the repeatability of its visibility is indicated on the timescale 
of $\approx 20$ years (Sect.~\ref{ss:0527}). 

According to the above-mentioned reasons we suggest that the 
brightest SSSs could be unidentified optical novae that are in 
a post-nova SSS state, when the residual surface nuclear burning 
is supported by rapid re-accretion from the donor, which can 
help in sustaining the high luminosity of the burning WD for 
a long time. 
The lifetime of the high luminosity, possibly super-Eddington, 
will then depend on the energy output from both the residual 
burning and the resumed accretion. 
%

%
\subsection{Mass-loss rate from the nebular emission}
\label{ss:dotM}
The high luminosities of SSSs, whose WDs probably accrete at 
a high rate, lead to a mass-outflow in the form of a wind 
\citep[e.g.,][and Appendix~\ref{s:appA} here]{1994ApJ...437..802K,
2001ApJ...558..323H}. 
According to \cite{1994ApJ...437..802K} the optically thick/thin 
interface of the wind represents the SSS pseudo-photosphere. 
In our cases, its average value is 
$R_{\rm SSS}^{\rm eff}\sim 0.17$\ro\ (Table~\ref{tab:par}). 
The optically thin wind above the hot and luminous pseudo-photosphere 
is ionized, giving rise to the nebular radiation 
(see Appendix~\ref{s:appA}). 
Assuming that the wind flows out spherically symmetrically 
at a constant velocity $v_{\infty}$, begins at the 
$R_{\rm SSS}^{\rm eff}$, and its radial density distribution 
satisfies the mass continuity equation, then the relationship 
between the mass-loss rate of the ionized wind, 
$\dot M_{\rm SSS}$, and its emission measure, $EM$, can be 
expressed as, 
%
\begin{equation}
 \dot M_{\rm SSS} = 
     \left[4\pi\,(\mu m_{\rm H} v_{\infty})^2
     R_{\rm SSS}^{\rm eff} EM\right]^{1/2}  {\rm g s^{-1}},
\label{eq:mdot1}
\end{equation}
\citep[see Eq.~(7) of][]{2014A&A...569A.112S}, where $\mu$ is 
the mean molecular weight and $m_{\rm H}$ is the mass of 
the hydrogen atom. This relation assumes that the outer radius 
of the ionized wind $\gg R_{\rm SSS}^{\rm eff}$. 
The average value of $R_{\rm SSS}^{\rm eff} = 0.17$\ro\ and 
$EM\gtrsim 10^{60}$\cmt\ (Table~\ref{tab:par}) yield 
$\dot{M}_{\rm SSS}\gtrsim 2\times 10^{-6}$\myr\ 
for $v_{\infty}\equiv 1000$\kms. 

Using the same assumptions as for the nebular continuum, 
similar values of $\dot M_{\rm SSS}$ can be obtained also from 
hydrogen emission lines, because they are formed in the same 
volume of the ionized hydrogen as the continuum. For example, 
for the measured flux in the \hb\ line, $F_{{\rm H}\beta}$, 
we can express its luminosity as, 
\begin{equation}
L_{{\rm H}\beta} = 4\pi d^2 F_{{\rm H}\beta} = 
                   h\nu_{\beta}\alpha({\rm H}\beta,T_{\rm e})
                   \int_{R_{\rm SSS}^{\rm eff}}^{\infty}\!\! n_{\rm e}(r)
                   n_{\rm p}(r)\,{\rm d}V, 
\label{eq:lhb}
\end{equation}
where $\alpha({\rm H}\beta,T_{\rm e})$ is the effective 
recombination coefficient for the \hb\ transition, and $n_{\rm e}$ 
and $n_{\rm p}$ is the concentration of electrons and protons in 
the spherical H$^{+}$ zone with the inner radius 
$R_{\rm SSS}^{\rm eff}$, from which the wind can become optically 
thin in the \hb\ line. The integral in Eq.~(\ref{eq:lhb}) represents 
the emission measure of the H$^{+}$ region (see Sect.~\ref{ss:sed}). 
Using its expression from Eq.~(\ref{eq:mdot1}), we obtain, 
%
\begin{equation}
 \dot M_{\rm SSS} = \left[4\pi\,(\mu m_{\rm H} v_{\infty})^2
                    R_{\rm SSS}^{\rm eff}\frac{L_{{\rm H}\beta}}
                    {h\nu_{\beta}\alpha({\rm H}\beta,T_{\rm e})}
                    \right]^{1/2}  {\rm g s^{-1}} ,
\label{eq:mdot2}
\end{equation}
which corresponds to a lower limit of $\dot M_{\rm SSS}$, because 
in a line transition the wind becomes optically thin above 
the photosphere, at distances $>R_{\rm SSS}^{\rm eff}$. This 
problem was treated by \cite{1988ApJ...326..356L}, who found 
that stellar winds of luminous hot stars become optically thin 
in the \ha\ line from a distance of $\sim$1.5 times the star's 
radius. 
Accordingly, $F_{{\rm H}\beta} = 1.45\times 10^{-13}$\ecs\ 
measured in the spectrum of LIN~333 by \cite{1987ApJ...320..159A} 
corresponds to $\dot{M}_{\rm SSS}\sim 1.7\times 10^{-6}$\myr\ 
for 
$\alpha({\rm H}\beta,30000) = 1.088\times 10^{-14}$\,cm$^3$\,s$^{-1}$ 
\citep[][]{1987MNRAS.224..801H}, $R_{\rm SSS}^{\rm eff} = 0.17$\ro\ 
and $v_{\infty}\equiv 1000$\kms. 
These observational values of $\dot{M}_{\rm SSS}$ are also 
consistent with those suggested by the optically thick wind 
theory, for winds from the nuclear-burning WDs of a high mass 
\citep[see Eq.~(23) of][]{1994ApJ...437..802K}. 
It is of interest to note 
that wind mass-loss rate of the order of $10^{-6}$\myr\ was 
measured during the long-lasting SSS-phase of the classical 
nova V339~Del, whose bolometric luminosity was kept at a high 
level of $1-2\times 10^{39}$\es\ for $\sim$150 days 
\citep[see Fig.~12 of][]{2019ApJ...878...28S}. 

\subsection{Dense unresolved circumstellar nebulae in SSSs}
\label{ss:ducn}

According to the mass continuity equation the particle density 
of the wind $n(r)\propto r^{-2}$, where $r$ is the radial distance 
from the source of the wind. 
For example, $n(1\,R_{\odot})\sim 9\times 10^{12}$\cmt, while 
$n(1\,AU)\sim 2\times 10^{8}$\cmt\ for wind parameters 
given in Sect.~\ref{ss:dotM}. Further, the emitted flux by 
the wind at the distance $r$ is proportional to $n^2$ (e.g., 
Eq.~(\ref{eq:lhb})), so the densities in our example correspond 
to the flux ratio of the wind layers, $F(1\,AU)/F(1\,R_{\odot}) = 
4\pi(1\,AU)^2n(1\,AU)^2/4\pi(1\,R_{\odot})^2n(1\,R_{\odot})^2
\sim 2\times10^{-5}$. This implies that the vast majority of 
the nebular emission is produced within a relatively small 
central part of the ionized wind -- from its optically thick/thin 
interface up to a few AU. 
Such dimensions of nebular medium are similar to those generated 
by accreting nuclear-burning WDs in symbiotic stars 
\citep[][]{2005A&A...440..995S}, and were also considered by 
\cite{2015MNRAS.453.2927N} in calculations the circumstellar 
mass-loss rates required for obscuration of supersoft X-ray 
sources. 
For the purpose of our work, it is important to note that such 
a small size of the circumstellar nebulae in SSSs at the distance 
of LMC and SMC is not spatially resolvable by the current 
observational technique. 
Further, the circumstellar nebulae generated by the ionized wind 
in a binary are also very dense. In our cases, the emission 
measure ($\sim n^2 V$) of a few times $10^{60}$\cmt\ 
(Table~\ref{tab:par}) corresponds to the mean particle density 
of a few times 10$^{9}$\cmt. 
Such the dense nebulae are characterized with a specific type of 
the emission line spectrum, whose fluxes and their theoretical 
ratios significantly differ from those produced by the low-density 
planetary nebulae and/or ISM nebulae predicted to be photoionized 
by the central SSS (see suggestions for future work in 
Sect.~\ref{s:sum}). 
Here, for a comparison with a dense circumstellar nebula in the 
symbiotic star AX~Per, \cite{2001A&A...367..199S} found the electron 
concentration in the H\,{\scriptsize II} and [\oiii] zone 
surrounding its burning WD to be of a few times $10^9$ and 
$\approx 3\times10^7$\cmt, respectively. Further, they suggested 
that the extremely steep Balmer decrement (e.g., \ha/\hb\ $\sim 7-10$) 
could be due to electron collisions in a very dense 
H\,{\scriptsize II} region. 

Finally, we note that the dense unresolvable circumstellar nebulae 
can be indicated for a given object only by the method of 
disentangling its composite spectrum (Sect.~\ref{ss:sed}). 

\subsection{Origin of the X-ray/optical flux anticorrelation}
\label{ss:anti}
The presence of a strong nebular radiation in the UV to NIR 
spectrum of our targets ($EM \gtrsim 2\times 10^{60}$\cmt, 
Table~\ref{tab:par}) and its variation 
(Fig.~\ref{fig:rx0513sed2}, Sect.~\ref{ss:0513sed}) allow 
us to explain the anticorrelation between their supersoft X-ray 
and optical fluxes by a variable mass-outflow from the accreting 
burning WD as in the case of the symbiotic binary AG~Dra 
\citep[see][]{2008A&A...481..725G,
              2009A&A...507.1531S}. 
It is assumed that the variable mass-outflow results from a variable 
mass-transfer. \cite{2010NewAR..54...75C} suggested a similar view 
for the SSS RX\,J0513.9-6951, and also estimated the mass outflow 
rate of $\approx 10^{-7}$\myr\ for the luminosity at the Eddington 
limit. 

An increase of the mass-loss increases the particle density above 
the hot WD's pseudo-photosphere, and thus the number of {\rm b--f} 
absorptions. This leads to a {\em decrease} of the supersoft 
X-ray photons, and consequently to an increase of {\rm f--b} and 
{\rm f--f} transitions that causes an {\em increase} of the nebular 
emission, which dominates the spectrum from the near-UV to longer 
wavelengths. 

In this way the enhanced (wind) mass-outflow increases the optical 
depth for the supersoft X-ray photons, and thus increases its 
optically thick/thin interface, which is the WD's pseudo-photosphere. 
That is why we indicate a larger effective radius of the SSS 
during the X-ray-off/optical-high states. 
For RX\,J0513.9-6951 the $R_{\rm SSS}^{\rm eff}$ radius inflated 
from $\sim$0.19\ro\ during the optical-low state to $\sim$0.26\ro\ 
during the optical-high state (Sect.~\ref{ss:0513sed}, 
Table~\ref{tab:par}). 
Independently, the connection between the flux anticorrelation 
and the WD's radius was demonstrated for RX\,J0513.9-6951 by 
\cite{2010NewAR..54...75C} on the basis of the \textsl{XMM-Newton} 
simultaneous soft X-ray, UV and optical observations (see their 
Figs.~2 and 3). 
Finally, the inflated WD can also increase its radiation for 
$\lambda > 912$\,\AA\ (compare the blue lines in the panel 
{\bf a} and {\bf b} of Fig.~\ref{fig:rx0513sed2}). 

In this way, the ionization/recombination process in the 
{\em variable} wind from the SSS causes the simultaneous 
presence of the X-ray-on/off and optical-low/high states. 
Corresponding changes in the nebular emission can explain 
$\sim$1\,mag changes in the optical brightness 
(Sect.~\ref{ss:0513sed}). 

\section{Conclusions and future work}
\label{s:sum}
In this paper we performed the multiwavelength modeling of 
the supersoft X-ray---NIR SED for the brightest Magellanic 
Cloud SSSs, RX\,J0513.9-6951, RX\,J0543.6-6822, 
RX\,J0058.6-7135 and RX\,J0527.8-6954. 
The global SED models satisfactorily fit the measured multi-band 
fluxes. 
The main results of the modeling can be summarized as follows. 
\begin{enumerate}
\item
The SED models revealed that their fundamental parameters, 
$L_{\rm SSS}$, $T_{\rm BB}$ and $R_{\rm SSS}^{\rm eff}$ are 
$\gtrsim 10^{38}-10^{39}$\es, $\approx3\times 10^{5}$\,K and 
$\sim0.17$\ro, respectively. 
Their supersoft radiation is attenuated with the total 
$N_{\rm H}\sim (7 - 12)\times 10^{20}$\cmd. 
The modeling identified the nebular component of radiation 
with $EM\gtrsim 2\times 10^{60}$\cmt\ (Table~\ref{tab:par}). 
Luminosities of our targets differ significantly from those 
inferred previously from modeling the X-ray data only 
(Sect.~\ref{ss:comp}, Fig.~\ref{fig:comparison}). 
\item
In spite of super-Eddington luminosities for a 1.4\mo\ compact 
object, the accretor has to be a WD, because the X-ray spectrum 
profile and a low speed of jets are not consistent with those of 
accreting NS or BH (Sects.~\ref{sss:ns} and \ref{sss:bh}). 
\item 
The high luminosities lead to the mass-loss at high rates. 
Assuming that the nebular component of radiation is produced by 
the ionized wind from SSSs, its emission measure corresponds to 
the mass-loss rate $\gtrsim 2\times 10^{-6}$\myr\ 
(Sect.~\ref{ss:dotM}). The ionized wind represents a dense 
unresolved circumstellar nabula in SSSs (Sect.~\ref{ss:ducn}, 
Appendix~\ref{s:appA}). 
\item
Accordingly, we suggest that the brightest SSSs in LMC and SMC 
could be unidentified optical novae in a post-nova SSS state, 
whose lifetime and luminosity are enhanced by the resumed 
accretion, possibly at super-Eddington rates 
(Sect.~\ref{sss:postnova}). 
\item
The observed anticorrelation between the supersoft X-ray and 
optical fluxes can be caused by a variable mass-outflow from 
the SSS due to variations in the mass-transfer. A higher 
mass-loss absorbs more X-ray photons that are re-emitted 
in the form of the nebular emission, and vice versa. 
As a result, we observe simultaneously the X-ray-off/on and 
optical-high/low states (Sect.~\ref{ss:anti}). 
\end{enumerate}

For future investigation, we suggest to use the multiwavelength 
modeling including simultaneously obtained fluxes from both 
ascending and descending part of the SSS spectrum. 
In this way to determine more reliable parameters than can be 
obtained by modeling only one part of the spectrum. 
To test the proposed model, and develop its more perfect 
version we suggest to acquire new observations, and using 
better theoretical modeling. In particular the following. 
\begin{itemize}
\item
Flux calibrated low-resolution optical/NIR spectrum 
($\lambda\gtrsim 350$\,nm) should verify the presence of 
the nebular continuum in the spectrum by its specific profile 
and the presence of the Balmer jump in emission. Moreover, for 
LIN~333, such the spectrum should confirm or refute its binary 
nature proposed in Appendix~\ref{s:appC}. 
\item
High-resolution spectra should directly indicate a high-velocity 
mass-outflow by (presumably) broad wings of the strongest lines 
as \ha, \hb\ and \heii\,$\lambda$4686. 
In addition, for RX\,J0513.9-6951, the presence of satellite 
components to these lines would indicate a collimated bipolar 
outflow resulting from a very high accretion rate onto the WD 
\citep[see][]{1996ApJ...470.1065S}. 
\item
In accordance with the study of \cite{2015MNRAS.453.2927N}, using 
the photoionization code {\small CLOUDY}, determine the critical 
value of the mass-loss rate, $\dot M_{\rm crit}$, from the burning 
WD that causes the obscuration of the central X-ray source for 
parameters determined by our modeling. 
The corresponding $EM$ should be comparable with that obtained 
from the SED modeling during the X-ray-off states. 
In this way to test the origin of the X-ray/optical flux 
anticorrelation by a variable wind from the SSS proposed 
in Sect.~\ref{ss:anti}. 

Furthermore, the calculated emission line spectrum produced 
by the ionized wind from the X-ray source should provide 
additional constraints on the SSS environment. 
By analogy with symbiotic stars, a dense (unresolved) nebula, 
here suggested also for SSSs by our SED modeling 
(Sect.~\ref{ss:ducn}; Appendix~\ref{s:appA}), should be 
characterized by an anomalous Balmer decrement 
(e.g., \ha/\hb$\gg 3$), and a very low rate of the nebular 
[\oiii] lines to [\oiii]\,$\lambda$4363 -- in strong contrast 
to values produced by the low-density planetary nebulae 
\citep[e.g.,][]{1997pdpn.book.....G}, and/or nebulae predicted 
to exist around SSSs imbedded in the low-density ISM 
\citep[][Sect.~\ref{ss:ducn} here]{2015MNRAS.453.2927N}. 
\item
In modeling the global SED, consider also the component of 
radiation from the irradiated accretion disk proposed by 
\cite{1996LNP...472...65P}. Its possible contribution in 
the far-UV will reduce the radiation from the burning WD 
here, and thus its bolometric luminosity, and amount of 
the required absorption. 
\end{itemize}

These tasks present new challenges for further observations 
and theoretical modeling, which should help us in understanding 
the nature and evolutionary status of the most luminous SSSs. 

\begin{acknowledgments}
The author thanks the anonymous referee for constructive comments. 
This research is based on observations made with the Far Ultraviolet 
Spectroscopic Explorer, International Ultraviolet Explorer obtained 
from the MAST data archive at the Space Telescope Science Institute, 
which is operated by the Association of Universities for Research 
in Astronomy, Inc., under NASA contract NAS 5–26555, and 
on observations made with the NASA/ESA Hubble Space Telescope, 
and obtained from the Hubble Legacy Archive, which is a collaboration 
between the Space Telescope Science Institute (STScI/NASA), 
the Space Telescope European Coordinating Facility (ST-ECF/ESAC/ESA) 
and the Canadian Astronomy Data Centre (CADC/NRC/CSA). 
\textsl{DENIS} is the result of a joint effort involving human 
and financial contributions of several Institutes mostly
located in Europe. It has been supported financially mainly
by the French Institut National des Sciences de l'Univers,
CNRS, and French Education Ministry, the European Southern   
Observatory, the State of Baden-Wuerttemberg, and the European
Commission under networks of the SCIENCE and Human Capital
and Mobility programs, the Landessternwarte, Heidelberg and
Institut d'Astrophysique de Paris.
This publication makes use of data products from the Two Micron
All Sky Survey, which is a joint project of the University of   
Massachusetts and the Infrared Processing and Analysis Center/California
Institute of Technology, funded by the National Aeronautics and Space   
Administration and the National Science Foundation. 

This research was supported by the Slovak Research and 
Development Agency under the contracts No. APVV-15-0458 and 
APVV-20-0148, and by the Slovak Academy of Sciences grant 
VEGA No. 2/0030/21. 
\end{acknowledgments}
%
\facilities{ROSAT(PSPC), XMM-Newton, Chandra, FUSE, HST(GHRS, FOS), 
            IUE(SWP, LWP), USNO-B Catalog, NOMAD Catalog, 
            2MASS All-Sky Catalog, 
            IRSF Magellanic Clouds Point Source Catalog, 
            2MASS 6X Point Source Working Database/Catalog, 
            The SMC Stellar Catalog and Extinction Map, 
            The LMC Stellar Catalog and Extinction Map, 
            Catalogue of Stellar Spectral Classifications, 
            the 3rd release of the DENIS database}
%
%
%

\appendix 

\section{Justification of nebular continuum in the spectrum of 
         the supersoft X-ray sources}
\label{s:appA}
\begin{figure}
 \begin{center}
\resizebox{10cm}{!}{\includegraphics[angle=-90]{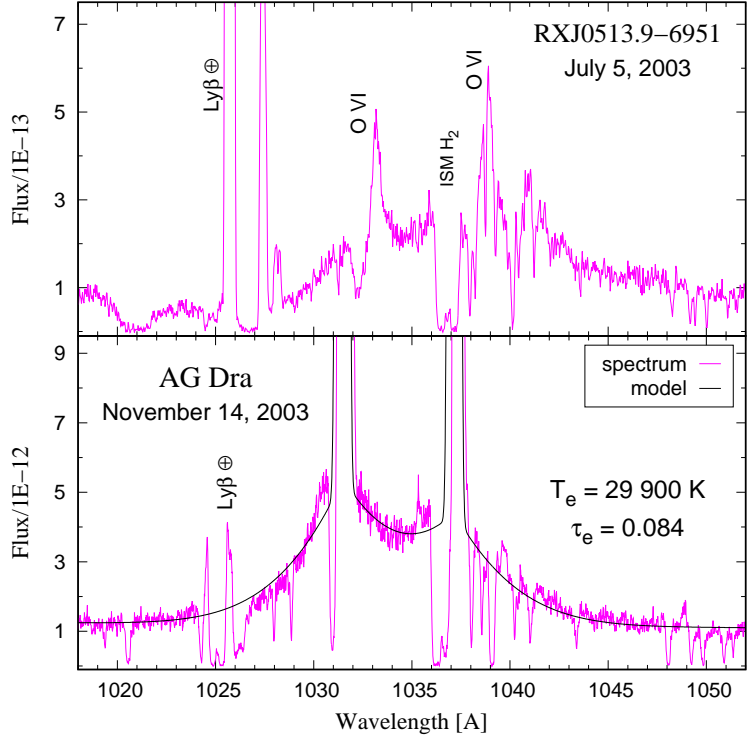}}
 \end{center}
\caption{
Broad wings of the resonance \ovi\,$\lambda$1032, $\lambda$1038 
doublet observed in the \textsl{FUSE} spectra of LMC SSS 
RX\,J0513.9-6951 (top panel) and of the X-ray symbiotic binary 
AG~Dra with the model of the scattering of line photons on free 
electrons \citep[bottom panel, adopted from][]{2012MNRAS.427..979S}. 
The similarity of these profiles suggests the presence of 
a dense circumstellar nebula also in RX\,J0513.9-6951, showing 
mass-outflow (see text). 
Note that the spectra of RX\,J0513.9-6951 and AG~Dra are shifted 
in wavelengths due to their space velocity by $\sim$+500 and 
$\sim$-150\kms, respectively 
\citep[][]{2005AJ....129.2792H,2000AJ....120.3255F}. 
}
\label{fig:thomson}
\end{figure}

%
The nebular component of radiation has not yet been considered 
in modeling the SED of SSSs. 
We mean the spatially unresolved source of nebular emission, the 
presence of which in the spectrum is indicated by multiwavelength 
modeling of the SED (Sect.~\ref{ss:sed} and Fig.~\ref{fig:seds}) 
as in the case of dense nebulae in symbiotic binaries and/or 
classical novae 
\citep[see][Sect.~\ref{ss:ducn} here]{2005A&A...440..995S,
                                      2015NewA...36..116S,
                                      2019ApJ...878...28S}. 
Its presence also in the spectrum of SSSs can be justified as 
follows. 
\begin{enumerate}
\item
It is probable that extremely luminous and hot SSSs will generate 
an outflow of material driven by radiation pressure as it is 
observed and theoretically justified for 
luminous hot stars 
\citep[e.g.,][]{1975ApJ...195..157C,
                1978ApJ...220..902K}, 
accreting WDs in symbiotic binaries 
\citep[e.g.,][]{1994A&A...284..145V,
                2006A&A...457.1003S,
                2017A&A...604A..48S}, 
classical novae 
\citep[e.g.,][]{1966MNRAS.132..317F,
                1976MNRAS.175..305B,
                1994ApJ...437..802K,
                2019ApJ...878...28S}, 
and in SSSs 
\citep[e.g.,][]{1985SSRv...40..229P,
1993A&A...278L..39P,
2010NewAR..54...75C,
2010A&A...517L...5O}. 
Simultaneously, the hot and luminous central WD and the boundary 
layer of accretion disk generate a large flux of hydrogen ionizing 
photons ($\sim 10^{48}$\,s$^{-1}$ for $T_{\rm BB} \sim 300\,000$\,K 
and $L_{\rm SSS} \sim 10^{38}$\es), which ionizes the out-flowing 
material giving rise to nebular emission. 
This reprocessed radiation dominates the Rayleigh-Jeans tail of 
a hot source radiating at $T_{\rm BB}\gtrsim 10^5$\,K, usually 
from the near-UV to longer wavelengths 
\citep[see][]{2005A&A...440..995S,2015MNRAS.453.2927N}. 
This nebular component of radiation is aimed on explaining 
the strong UV--NIR excess observed for bright SSSs. 
%
\item
Below we summarize indications of the nebular emission 
observed in the spectrum of our targets. 
\begin{itemize}
\item
Mass-loss from the accreting nuclear-burning WD is indicated by 
P-Cyg profiles of H lines, their often strong flux and broad 
emission wings, and/or satellite components to main \hb\ and 
\heii\,$\lambda$4686 emission cores as for RX\,J0513.9-6951 
(see references above and those in 
Sects.~\ref{ss:0513} to \ref{ss:0527}). 
Therefore, the \ha\ line flux as a tracer of mass loss 
from hot stars has been usually used to estimate its rate 
\citep[e.g.,][]{1978ApJ...220..902K,
                1988ApJ...326..356L,
                1999isw..book.....L,
                2002MNRAS.335.1109S,
                2006A&A...457.1003S}. 
The out-flowing material ionized by the central hot source 
then gives rise to nebular emission. 
\item 
During the optical high state of RX\,J0513.9-6951, the resonance 
\ovi\,$\lambda$1032, $\lambda$1038 doublet shows broad wings 
due to electron scattering, with the superposed central P~Cyg-type 
of the profile \citep[see][]{2005AJ....129.2792H}. 
By comparison with accreting nuclear-burning WDs in symbiotic stars, 
such the wings are created in the layer of electrons, throughout 
which the line photons are transferred, with the electron optical 
depth $\tau_{\rm e}\sim0.05 - 0.6$, depending on the system and 
its activity \citep[see][and Fig.~\ref{fig:thomson} here for 
illustration]{2012MNRAS.427..979S}. As there is no resolved 
ionization nebula surrounding RX\,J0513.9-6951, the layer of free 
electrons has to be spread in the vicinity of the ionizing source, 
which indicates the presence of a compact (unresolved) circumstellar 
nebula in RX\,J0513.9-6951, independently to our SED model. 
For a comparison, in the case of the X-ray symbiotic binary AG~Dra 
the layer of scattering electrons extends to $\approx$10 -- 100\ro, 
which corresponds to the mean electron concentrations of 
$10^{10} - 10^{12}$\cmt\ to satisfy the measured $\tau_{\rm e}$ 
\citep[see][]{2009A&A...507.1531S}. 
\item
The presence of emission lines directly indicates the presence 
of the corresponding nebular continuum, because both continuum 
and lines originate in the same ionized volume. 
Example here is the case of RX\,J0058.6-7135 (LIN 333), 
for which \cite{1987ApJ...320..159A} revealed the presence of 
a strong nebular emission radiating at a high 
$T_{\rm e}\sim 30000$\,K. 
In particular, the measured (extinction free) flux in the \hb\ 
line, $F_{{\rm H}\beta} = 1.45\times 10^{-13}$\ecs\ corresponds 
to the emission measure, 
\begin{equation}
  \textsl{EM} = 4\pi d^2 \frac{F_{{\rm H}\beta}}
                {h\nu_{\beta}\,\alpha({\rm H}\beta)}
 = 1.4\times 10^{60}\,(d/60\,{\rm kpc})^2\,{\rm cm}^{-3},
\label{eq:emhb}
\end{equation}
which agrees well with our value determined from the nebular 
continuum (see Table~\ref{tab:par})\footnote{Note that values 
of parameters derived from the emission lines are usually 
smaller than those derived from the continuum, because 
nebulae are usually more opaque in the lines than in 
the continuum.}. 
In Eq.~(\ref{eq:emhb}), we used an effective recombination 
coefficient for the \hb\ transition and $T_{\rm e} = 30000$\,K, 
$\alpha({\rm H}\beta) = 1.088\times 10^{-14}$\,cm$^3$\,s$^{-1}$ 
\citep[][]{1987MNRAS.224..801H}. 
\item
Often measured flux ratio F(\heii\,$\lambda$4686)/F(\hb)$\gg$1 suggests 
the presence of the nebular continuum from doubly ionized helium, we 
can indicate as a discontinuity at $\sim 2050$\,\AA\ (see 
Figs.~\ref{fig:seds}c, \ref{fig:rx0513sed2}a and \ref{fig:CAL83app}). 
\item
In spite that the He$^{++}$ region around strong SSSs should 
be especially prominent as compared to other astrophysical nebulae 
\citep[][]{1994ApJ...431..237R}, its presence within the resolved 
ionization nebula around CAL~83 had not been detected for a long 
time \citep[][]{1995ApJ...439..646R}. 
The first evidence of an \heii$\lambda$4686 region around CAL~83 
was presented by \cite{2012A&A...544A..86G} as a faint 
(relative to its bright stellar component) asymmetrically 
distributed zone confined to one side of CAL~83. In addition, 
the authors found that the \hii\ emission did not fit in with 
model predictions. 
This unusual \heii$\lambda$4686 nebular emission could be 
explained by the presence of a compact (unresolved) circumstellar 
nebula around the burning WD of CAL~83, whose He$^{++}$ zone is 
almost ionization-bounded, i.e., almost all photons with 
$\lambda < 228$\,\AA\ are absorbed within this zone. 
%
For a comparison, in the case of the eclipsing symbiotic star 
AX~Per, the He$^{++}$ region extends up to $\sim$50\ro\ from 
the WD within its dense wind escaping at a rate of 
$\sim 2\times 10^{-6}$\myr\ and generating the total emission 
measure of a few times $10^{59}$\cmt\ 
\citep[from the eclipse profile, see][]{2011A&A...536A..27S}. 
\item
Similarly to the above point, the circumstellar nebulae in the 
brightest SSSs as suggested by our SED models could attenuate 
also the flux of H-ionizing photons to such an extent that it 
would not allow us to detect the corresponding extended ISM 
nebulae in the optical. 
From this point of view, the strange absence of ionization ISM 
nebulae surrounding the brightest SSSs in LMC and SMC, up to 
CAL~83 \citep[][]{1995ApJ...439..646R,
                  2016MNRAS.455.1770W,
                  2020MNRAS.497.3234F} 
could be explained by the presence of compact (unresolved) 
circumstellar nebulae in these objects. 
\end{itemize}
\end{enumerate}
%
%
%

\section{CAL~83: The SED of the central source and 
         the surrounding nebula}
\label{s:appB}
\begin{figure}
 \begin{center}
\resizebox{10cm}{!}{\includegraphics[angle=-90]{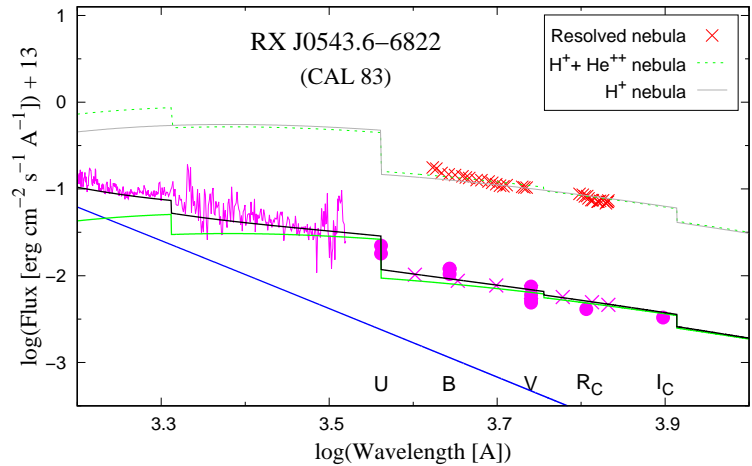}}
 \end{center}
\caption{
The UV--optical part of the CAL~83 SED from Fig.~\ref{fig:seds}c 
including representative flux-points from the resolved 
$7.5\times7.5$\,pc$^2$ nebula surrounding CAL~83 (red crosses), 
taken from Fig.~2 of \cite{2012A&A...544A..86G}. 
They can be matched by nebular continuum with the same $T_{\rm e}$ 
but significantly larger $EM$ (dotted green and gray line for 
H$^{+}$\,+\,He$^{++}$ and H$^{+}$ nebula) than the unresolved 
circumstellar nebula (solid green line; see text). 
Meaning of the other lines and symbols as in Fig.~\ref{fig:seds}. 
}
\label{fig:CAL83app}
\end{figure}

Figure~\ref{fig:CAL83app} shows the UV--optical part of the 
CAL~83 SED compared with that of the surrounding inner 
$7.5\times7.5$\,pc$^2$ ionization nebula. The fluxes of the 
latter (red crosses in the figure) were taken from Fig.~2 
of \cite{2012A&A...544A..86G}. 
The spectrum of the outer nebula can be matched with the same 
type of the nebular continuum (dotted green line) as the 
circumstellar nebula (solid green line), but scaled with a factor 
of $\sim$17 times larger. This suggests similar electron 
temperature ($\sim$30000\,K), but $\sim$17 times larger emission 
measure (i.e., $\sim$4.9$\times10^{61}$\cmt) in comparison with 
the dense circumstellar nebula
(see Table~\ref{tab:par}, Sect.~\ref{ss:ducn}). 

As noted above, there is a lack of doubly ionized helium in the 
surrounding resolved nebula (Appendix~\ref{s:appA}). Therefore, 
we also compared its fluxes with the net hydrogen nebular 
continuum that corresponds to $EM\sim 8.0\times10^{61}$\cmt\ and 
the same $T_{\rm e}$ (gray line in Fig.~\ref{fig:CAL83app}). 
Such the high $EM$ requires a high flux of the ionizing photons. 
Assuming that all H-ionizing photons from CAL~83 balances 
the rate of recombinations in both nebulae, we can estimate 
the corresponding $L_{\rm SSS}$ for the temperature of 
the ionizing source $T_{\rm BB}$ according to expression, 
\begin{equation}
  L_{\rm SSS} = \alpha_{\rm B}({\rm H},T_{\rm e})\,\textsl{EM}
               \frac{\sigma T_{\rm BB}^{4}}{f(T_{\rm BB})},   
\label{eq:lwdem}
\end{equation}
\citep[see][]{2017A&A...604A..48S}, 
where the function $f(T_{\rm BB})$ determines the flux of 
ionizing photons emitted by 1\,cm$^2$ area of the ionizing 
source, and $\alpha_{\rm B}({\rm H},T_{\rm e})$ is the 
recombination coefficient to all but the ground state of hydrogen 
(i.e., Case $B$). 
Then the total $EM\sim 8.29\times10^{61}$\cmt\ (i.e., of both 
nebulae), $T_{\rm BB} = 345000$\,K, 
$f(T_{\rm BB}) = 6.0\times 10^{27}$\,cm$^{-2}$\,s$^{-1}$, and 
$\alpha_{\rm B}({\rm H},30000) = 9.97\times10^{-14}$\cmtt\ 
\citep[][]{1987A&A...182...51N} correspond to 
$L_{\rm SSS}\sim 1.1\times10^{39}$\es. 
This value is close to that given by the SED models, but in 
the order of magnitude higher than those derived from the 
emission lines in previous works (see Sect.~\ref{ss:cal83}). 
Consequently, the very high $EM$ ($8.0\times10^{61}$\cmt) of 
the $7.5\times7.5$\,pc$^2$ nebula surrounding CAL~83 corresponds 
to its average particle density $\sim$39\cmt\ and the mass 
of $\sim$1700\mo, which are a factor of $>$4, and a factor of 
$\sim$11, respectively, larger than those previously derived 
from emission lines 
\citep[e.g.,][]{1995ApJ...439..646R,2012A&A...544A..86G}. 
Such the large difference of the surrounding nebula properties 
can be given by a large difference in the opacity of the emitting 
medium in the line and the continuum transitions, respectively. 
%
%
%

\section{LIN~333: Planetary nebula or a binary system?}
\label{s:appC}
%
Here we present two reasons suggesting that LIN~333 could be 
a binary system. 

\begin{figure}
 \begin{center}
\resizebox{\hsize}{!}
          {\includegraphics[angle=-90]{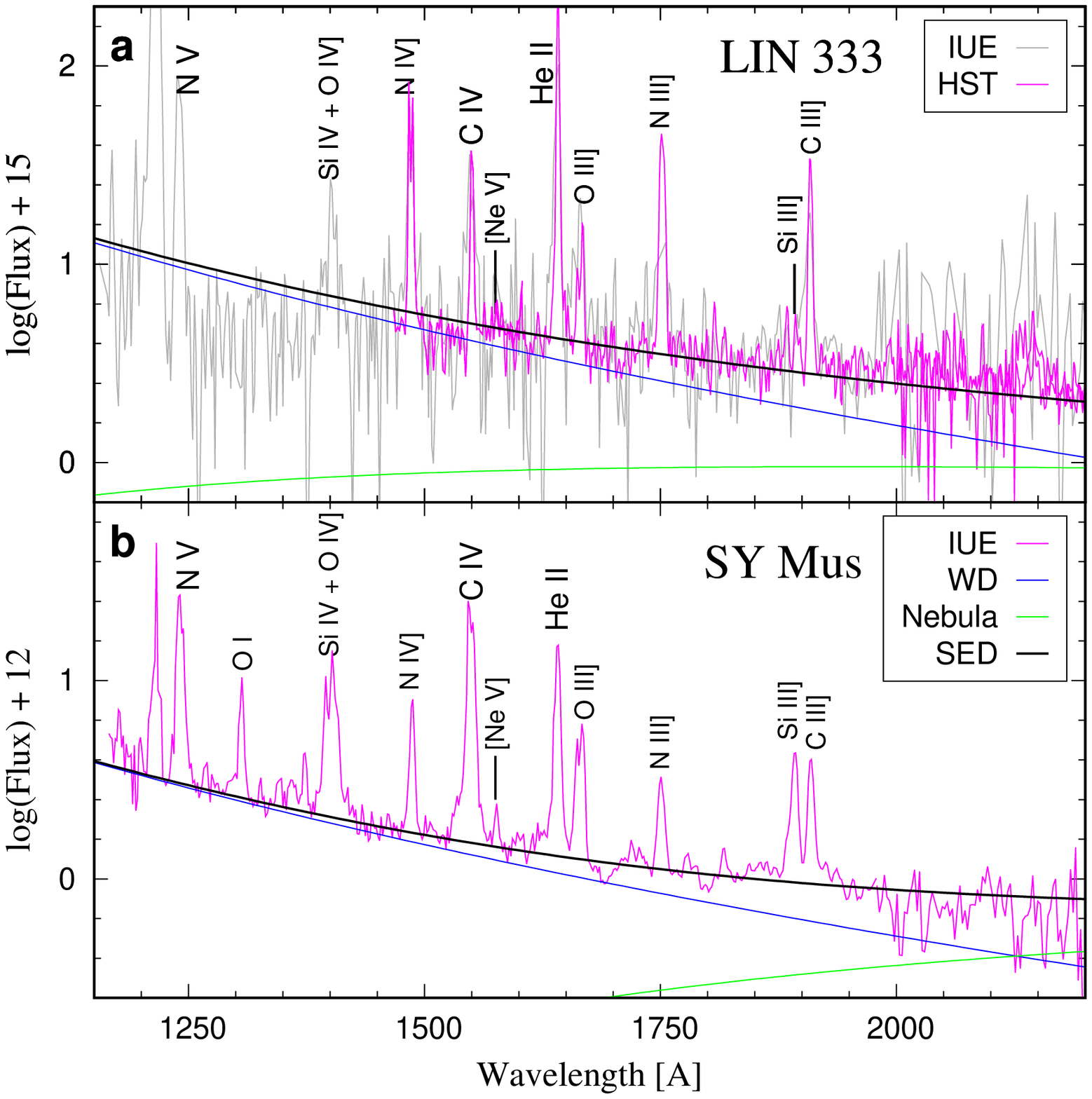}
           \includegraphics[angle=-90]{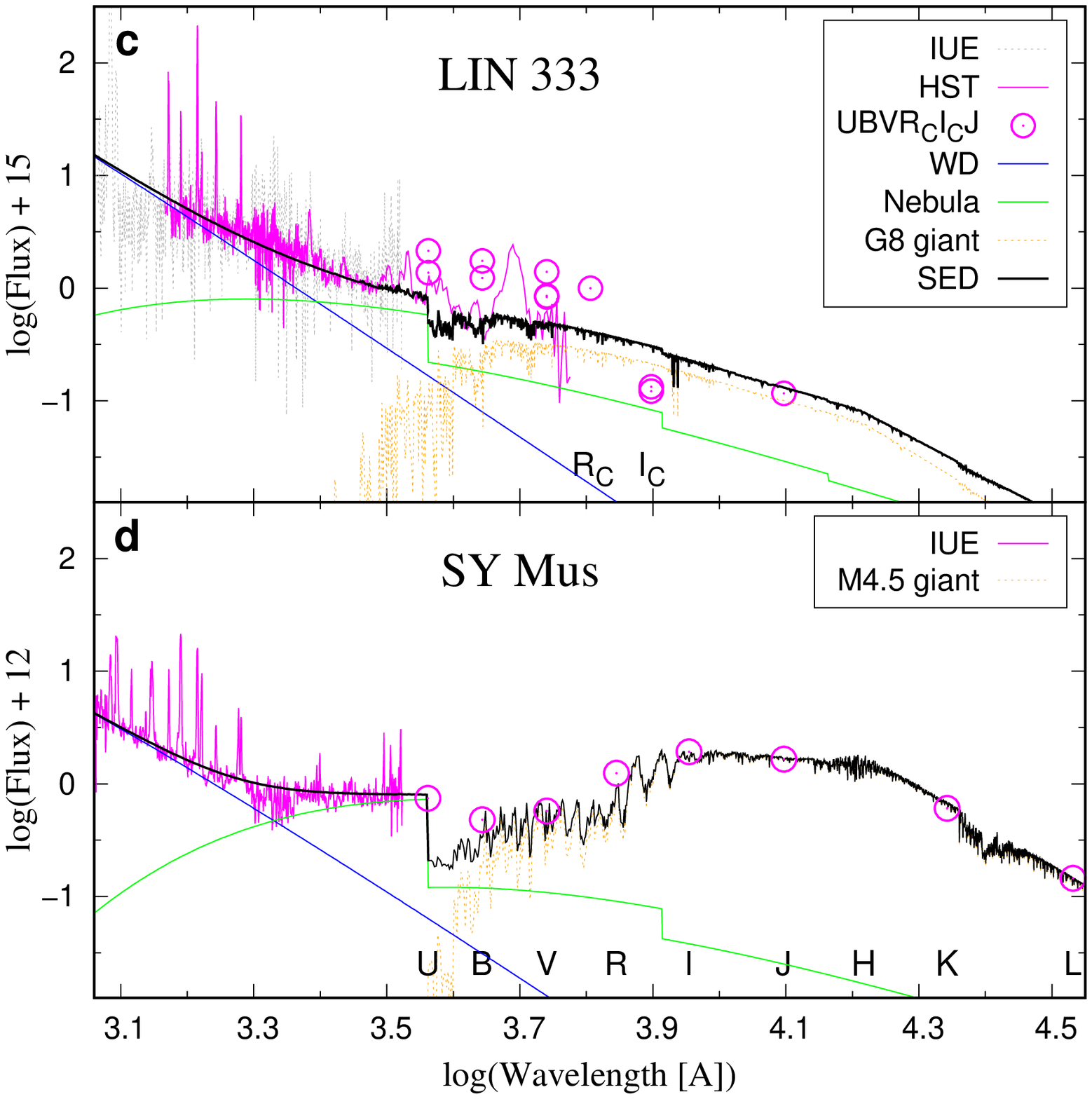}}
 \end{center}
\caption{
Left: Comparison of the far-UV spectrum emitted by the SSS LIN~333 
({\bf a}) and by the symbiotic binary SY~Mus ({\bf b}), dereddened 
with $E_{\rm B-V}$ = 0.08 and 0.35. The similarity of the nebular 
spectrum of both systems suggests their same origin -- in a compact 
circumstellar nebula (see text). 
Right: The UV--NIR SEDs of these objects. The case of LIN~333 allows 
the presence of a yellow giant companion (orange dashed line) to 
the SSS, suggesting binary nature of LIN~333 
(see Appendix~\ref{s:appC}). 
Panels {\bf a} and {\bf c} represent details of Fig.~\ref{fig:seds}b, 
while the panels {\bf b} and {\bf d} were adapted according to 
Fig.~18 of \cite{2005A&A...440..995S}. 
}
\label{fig:lin333app1}
\end{figure}

(i) 
First, we compare the nebular spectrum, both in the continuum 
and emission lines, of LIN~333 with that observed in symbiotic 
binaries containing an accreting nuclear-burning WD. Panels 
{\bf a} and {\bf b} of Fig.~\ref{fig:lin333app1} show a comparison 
with SY~Mus in the 1200 -- 2200\,\AA\ part of the spectrum. 
Both, LIN~333 and SY~Mus contain the same emission lines in 
the far-UV spectrum, except for the low-excitation 
O\,{\small I}\,$\lambda\lambda$1302-1306 lines\footnote{The 
absence of O\,{\small I} and very faint resonance 
\ovi\,$\lambda\lambda$1032,1038 doublet probably reflect 
a low abundance of oxygen in the LIN~333 nebula suggested 
by \cite{1996A&AS..116...95L}.}, which are characteristic 
for symbiotic stars. 
Strong resonance emission lines, 
N\,{\small V}\,$\lambda\lambda$1238,1242, 
C\,{\small IV}\,$\lambda\lambda$1548,1550, high-ionization line
He\,{\small II}\,$\lambda$1640, and the intersystem lines of 
O\,{\small IV}]\,$\lambda\lambda$1397-1405, 
N\,{\small IV}]\,$\lambda\lambda$1483-1486, 
O\,{\small III}]\,$\lambda\lambda$1661,1666, 
N\,{\small III}]\,$\lambda\lambda$1746-1753, 
Si\,{\small III}$\lambda$1892 and 
C\,{\small III}$\lambda\lambda$1907,1909 
reflect the circumstellar nebular conditions of symbiotic stars 
\citep[][]{1994ApJS...94..183M}, and thus also of SSS LIN~333. 

Also, the nebular continuum in the LIN~333 spectrum is 
unambiguously indicated by our method as for other targets 
(green lines in Fig.~\ref{fig:seds}). 
As there is no resolved ionization nebula around LIN~333, the 
indicated nebular continuum suggests the presence of a compact 
unresolved nebula in the system, the origin of which is similar 
to that of dense symbiotic nebulae, i.e., resulting from the 
interaction between the binary components, given by the ionized 
mass outflows 
\citep[e.g.,][]{1966SvA....10..331B,1984ApJ...284..202S,
                1987A&A...182...51N,2005A&A...440..995S}. 
%
Here, it is of interest to note that the first studies of 
symbiotic stars pointed to the similarity of their hot 
components to the central star of planetary nebulae 
\citep[e.g.,][]{1932PASP...44..318B}, and this possibility was 
often considered until the first satellite observations that 
unambiguously revealed the binary nature of symbiotic stars 
\citep[e.g.,][and references therein]{1984ApJ...279..252K}. 

Table~\ref{tab:appC} compares luminosities of the strongest 
emission lines in the UV spectrum, $EM$ and $T_{\rm e}$ 
of the compact nebulae of LIN~333 and SY~Mus. 
%
%
%
\begin{table*}
\caption{
Luminosities of the strongest permitted lines in the UV 
spectrum (in \es), emission measure $EM$ (\cmt), and 
electron temperature $T_{\rm e}$ (K) of LIN~333 and SY~Mus. 
The distance dependent parameters are scaled with 1.0\kpc\ 
\citep[][]{2005A&A...440..995S} and 60\kpc\ for SY~Mus 
and LIN~333, respectively. 
The corresponding spectra and models are shown in 
Fig.~\ref{fig:lin333app1}a and \ref{fig:lin333app1}b. 
}
\label{tab:appC}
\begin{center}
\begin{tabular}{cccccc}
\hline
\hline
\noalign{\smallskip}
Object                                 &
$L_{\rm N\,{\small V}\,\lambda1240}$   &
$L_{\rm C\,{\small IV}\,\lambda1550}$  &
$L_{\rm He\,{\small II}\,\lambda1640}$ &
$EM$                                   &
$T_{\rm e}$                            \\
%
%
\noalign{\smallskip}
\hline
\noalign{\smallskip}
LIN~333   & 2.7$\times10^{35}$ & 5.4$\times10^{34}$ & 2.9$\times10^{35}$ 
          & 1.6$\times10^{60}$ & $\sim$37000 \\
SY~Mus    & 2.1$\times10^{34}$ & 2.2$\times10^{34}$ & 1.1$\times10^{34}$ 
          & 3$\times10^{59}$   & $\sim$18500 \\
\noalign{\smallskip}
\hline
\end{tabular}
\end{center}
\normalsize
\end{table*}
%
%
%
%

(ii)
According to the $UBV$ photometry and the \textsl{HST} 
spectroscopy of LIN~333 (Sect.~\ref{ss:0058obs}), the measured 
fluxes in the visual band lie above our two-components model 
(see Fig.~\ref{fig:seds}b), which suggests contribution from 
the companion in a binary system. 
To match the optical continuum of the \textsl{HST} 
spectrum\footnote{We ignored the $U, B, V$ fluxes, because 
they are not corrected for emission lines, and thus lie well 
above the true continuum.} with a minimum effect on the 
near-UV region, where the original two-component model 
is satisfactory (see Fig.~\ref{fig:seds}b), we selected 
a synthetic spectrum of a yellow giant from a grid of models 
made by \cite{1999ApJ...525..871H}, calculated for 
$T_{\rm eff} = 5000$\,K, $\log g = 3.5$ and [M/H] = -1. 
Its scaling in Fig.~\ref{fig:lin333app1}c corresponds to the 
luminosity of 278\lo\ and the radius of 22\ro; parameters 
similar to a G8\,{\small III} giant 
\citep[e.g.,][]{1999AJ....117..521V}. If this were the case, 
LIN~333 would belong to the so-called yellow symbiotic stars. 
However, new optical/NIR observations are highly desirable 
to confirm or refute the binary nature of LIN~333. 
\clearpage
\bibliography{multiSSS}{}
\bibliographystyle{aasjournal}

\end{document}